\documentclass[aps,prb,twocolumn,superscriptaddress,showpacs]{revtex4}
\usepackage{graphicx,epsf}
\usepackage{amsmath,amssymb}
\usepackage{multirow}

\usepackage{color}
\newcommand{\w}{\omega}

\newcommand{\TK}{T_{\rm K}}

\newcommand{\tchi}{T\chi_{\rm imp}}
\newcommand{\s}{S_{\rm imp}}

\begin{document}
\title{
Quantum phase transitions and thermodynamics of the power-law Kondo model
}

\author{Andrew K. Mitchell}
\affiliation{Department of Chemistry, Physical and Theoretical Chemistry,
Oxford University, South Parks Road, Oxford OX1 3QZ, United Kingdom}
\affiliation{
Institut f\"ur Theoretische Physik, Universit\"at zu K\"oln,
Z\"ulpicher Stra\ss e 77, 50937 K\"oln, Germany
}
\author{Matthias Vojta}
\affiliation{Institut f\"ur Theoretische Physik, Technische Universit\"at Dresden,
01062 Dresden, Germany}
\author{Ralf Bulla}
\affiliation{
Institut f\"ur Theoretische Physik, Universit\"at zu K\"oln,
Z\"ulpicher Stra\ss e 77, 50937 K\"oln, Germany
}
\author{Lars Fritz}
\affiliation{
Institut f\"ur Theoretische Physik, Universit\"at zu K\"oln,
Z\"ulpicher Stra\ss e 77, 50937 K\"oln, Germany
}

\date{\today}

\begin{abstract}
We revisit the physics of a Kondo impurity coupled to a fermionic host with a diverging power-law
density of states near the Fermi level, $\rho(\w) \sim |\w|^r$, with exponent $-1<r<0$.
Using the analytical understanding of several fixed points, based partially on powerful
mappings between models with bath exponents $r$ and $(-r)$, combined with accurate
numerical renormalization group calculations, we determine thermodynamic quantities within
the stable phases, and also near the various quantum phase transitions.
Antiferromagnetic Kondo coupling leads to strong screening with a negative
zero-temperature impurity entropy, while ferromagnetic Kondo coupling can induce a stable
fractional spin moment. We formulate the quantum field theories for {\em all}
critical fixed points of the problem, which are fermionic in nature and allow for a
perturbative renormalization-group treatment.
\end{abstract}

\maketitle

%%%%%%%%%%%%%%%%%%%%%%%%%%%%%%%%%%%%%%%%%%%%%%%%%%%%%%%%%%%%%%%%%%%%%%%

\section{Introduction}

Impurities can act as probes for host materials. The Kondo effect,\cite{hewson} which describes the screening of a single magnetic impurity embedded in a metal at low temperatures, is a striking example: screening occurs (and can be probed locally), only when low-energy conduction electron-hole pairs are available.
Numerous generalizations of the Kondo effect have been discussed over the last two
decades, involving host systems such as superconductors,\cite{zitt70,sakaisc,withoff,cassa}
Luttinger liquids,\cite{fabrizio95,kim01}
boundary states of topological insulators,\cite{law10,zitko10,trankim10,akm:qpiti}
and spin liquids.\cite{ffv06,kim08,dhochak10,ribeiro12} In many of these cases, the interaction
between the impurity moment and the host excitations are non-trivial and lead to
zero-temperature phase transitions\cite{ssbook,mvrev} which can be tuned by varying impurity or host
parameters.

A prominent and rich example is given by the so-called pseudogap Kondo problem,
describing a magnetic moment embedded in a system of weakly interacting fermions with a
power-law vanishing density of states (DOS) $\rho(\w) \sim |\w|^r$ with $r>0$. Originally
discussed in the context of unconventional superconductors,\cite{withoff,cassa}
where $p$- and $d$-wave pairing symmetries yield $r=2$ and $1$ respectively,
the problem has continued to attract much attention; for example due to its
realization in semimetals like
graphene.\cite{baskaran07,wehling10,dellanna10,epl,uchoa11b,fvrop}
The pseudogap Kondo model displays various phases and quantum phase transitions (QPTs), which
have been studied extensively using both numerical and analytical
techniques.\cite{withoff,cassa,chen,bulla,ingersent,insi,GBI,mvrev,glossop11,vf04,fv04,DELpt,DELlma}
In particular, fermionic quantum field theories have been introduced and employed to
calculate perturbatively critical properties in the vicinity of certain critical
``dimensions'' --- i.e. special values of the bath exponent $r$.\cite{withoff,vf04,fv04}

The problem of a magnetic impurity coupled to fermions with a \emph{diverging} DOS, $\rho(\w)
\sim |\w|^r$ with $-1<r<0$ (hereafter referred to as the ``power-law Kondo model''),
can also be realized: A band edge in one space dimension leads to a DOS with $r=-\tfrac{1}{2}$,
disordered two-dimensional Dirac fermions can display a diverging DOS with varying
exponent, \cite{ludwig94,motru02}
and a diverging DOS might also appear at critical points within extensions\cite{edmft} of
dynamical mean-field theory.\cite{dmft}
The power-law Kondo model was investigated previously in Ref.~\onlinecite{vb02} using
Wilson's Numerical Renormalization group (NRG) technique,\cite{nrg,nrg_rev} with particular
emphasis on the structure of the phase diagram and the occurrence of a fractional-spin phase.
Moreover, a Kondo model with $r=-1^+$, i.e. a DOS of the limiting form $\rho(\w) \sim 1/(|\w|
\ln^2 |\w|)$, has been proposed for vacancy-induced moments in
graphene,\cite{cazalilla12} and properties of this model have recently been investigated
in detail using a combination of NRG and analytics in Ref.~\onlinecite{mf13}.

In this paper, we revisit the rich physics of the power-law Kondo model over the entire
bath exponent range $-1<r<0$, for two reasons: (i) While the initial study\cite{vb02} was
based mainly on NRG results, subsequent work on the positive-$r$
case\cite{vf04,fv04,serge05} has lead to substantial analytical insight into the
structure of both the stable and quantum critical fixed points of the Kondo model with
$|\w|^r$ bath DOS. Here we extend and apply these analytical concepts to the negative-$r$
case. (ii) The physical observables obtained in Ref.~\onlinecite{vb02} were rather
limited, and the limiting low-temperature properties were not correctly
obtained.\cite{error02} Here we present new and highly-accurate numerical results for the
full temperature-dependence of the impurity entropy and susceptibility as well as for
critical exponents, complemented and confirmed by analytical results, over the entire
range $-1<r<0$.

As the main result, we are able to provide and analyze the critical field theories for
all four intermediate-coupling fixed points of the problem. As in earlier work on
related models, these field theories are not of Landau-Ginzburg-Wilson type, but instead
are fermionic in nature. Using epsilon-expansion techniques, we calculate critical
properties which we find in perfect agreement with numerical results.

The body of the paper is organized as follows:
In Sec.~\ref{sec:model} we specify the power-law Kondo model
and introduce the observables to be discussed throughout the paper.
The phase diagram of the model is reviewed in Sec.~\ref{sec:phd}, supported and justified
by numerical results.
Sec.~\ref{sec:trivfp} is devoted to a discussion of the `trivial' (weak-coupling and
strong-coupling) fixed points of the model. Due to the power-law host DOS, most of these
fixed points nevertheless display somewhat non-trivial thermodynamic properties, which we analyze
using suitable effective models and mappings.
These are the basis for the construction of critical field theories for the various
\emph{intermediate}-coupling fixed points, which are subject of Sec.~\ref{sec:critfp}. We sketch
the renormalization group (RG) treatment of the field theories and calculate critical
properties.
Throughout the paper, analytical results are compared with highly accurate numerical
results obtained by NRG. A brief discussion will close the paper.

%%%%%%%%%%%%%%%%%%%%%%%%%%%%%%%%%%%%%%%%%%%%%%%%%%%%%
%%%%%%%%%%%%%%%%%%%%%%%%%%%%%%%%%%%%%%%%%%%%%%%%%%%%%
%%%%%%%%%%%%%%%%%%%%%%%%%%%%%%%%%%%%%%%%%%%%%%%%%%%%%

\section{Model and observables}
\label{sec:model}

%%%%%%%%%%%%%%%%%%%%%%%%%%%%%%%%%%%%%%%%%%%%%%%%%%%%%

\subsection{Power-law Kondo model}

The Kondo Hamiltonian for a spin-$\tfrac{1}{2}$ impurity can be written as
$H=H_b+H_{\rm K}$, with the impurity part
\begin{equation}
\label{hk}
H_{\rm K} = J \vec{S} \cdot {\vec s}_0 + V \sum_{\sigma}c_{0\sigma}^{\dagger} c^{\phantom{\dagger}}_{0\sigma}
\end{equation}
and the non-interacting conduction band described by
$H_{ b} = \sum_{{\vec k}\alpha} \epsilon^{\phantom{\dagger}}_{\vec k} c^\dagger_{{\vec k}\alpha} c^{\phantom{\dagger}}_{{\vec k}\alpha}$. Here ${\vec S}$ denotes a spin-$\tfrac{1}{2}$ operator for the impurity, and $\vec{s}_0 = \sum_{\alpha\beta} c^\dagger_{0\alpha} {\vec\sigma}_{\alpha\beta}
c^{\phantom{\dagger}}_{0\beta}$ is the conduction electron spin density at the impurity
site ${\vec r}_0 = 0$, with $c^{\phantom{\dagger}}_{0\sigma}=N_{\text{orb}}^{-1/2}\sum_{\vec{k}}c_{{\vec k}\sigma}$ and $N_{\text{orb}}\rightarrow \infty$.

For most of the discussion, we employ a conduction band with
a symmetric density of states and pure power law behavior,
\begin{equation}
\label{dos}
\rho(\w) = \rho_0 \left|\frac{\w}{D}\right|^r~~{\rm for}~|\w| < D = 1,
\end{equation}
where $\rho_0 = (1+r)/(2D)$. Particle-hole (p-h) asymmetry is
introduced via a potential scattering term $V\neq 0$ in Eq.~(\ref{hk}), see also Sec.~\ref{sec:asym} below.

Importantly, many properties discussed in this paper, such as critical exponents and
fixed-point observables, display universality in the sense that they are only determined
by the behavior of $\rho(\w)$ at {\em low} energies (assumed to be of the quoted
power-law form) and the presence or absence of p-h symmetry.

%%%%%%%%%%%%%%%%%%%%%%%%%%%%%%%%%%%%%%%%%%%%%%%%%%%%%

\subsection{Particle--hole asymmetry}
\label{sec:asym}

As p-h asymmetry plays a singular role for the power-law Kondo model at weak coupling,
it is worth reviewing some of its properties.
In a fermionic quantum impurity model, there are various sources of p-h asymmetry:
(i) The conduction band DOS can be asymmetric, $\rho(\w)\neq\rho(-\w)$. Here we shall
only consider a high-energy asymmetry, such that the low-energy DOS is symmetric:
$\lim_{|\w|\to 0} \rho(\w) = a_\pm |\w/D|^r$ for $\w\gtrless 0$ with $a_+=a_-$. This is the
case relevant for graphene, unconventional superconductors etc.
(ii) The impurity can induce potential scattering at the local bath site to which it is coupled.
In Eq.~\eqref{hk}, this corresponds to $V\neq 0$.
(iii) The impurity orbital itself can be p-h asymmetric, if described by an Anderson
impurity model\cite{hewson} (such asymmetry is obviously absent by construction in a Kondo model).

These sources of p-h asymmetry are usually considered to be qualitatively equivalent for
the low-energy behavior, since they all influence the real part of the free bath Green function
at the impurity site. Explicit links can be made as follows.

First, the equivalence of (iii) and (ii) can be established in the singly-occupied Kondo limit of an
Anderson impurity model via the Schrieffer-Wolff transformation.\cite{hewson} We note that this
requires finite Coulomb interaction in the Anderson model, and does not hold for a
non-interacting resonant level model.

Second, the equivalence of (i) and (iii) can be directly seen for an Anderson model,
where a real part of the bath Green function simply renormalizes the impurity level position.

Third, the equivalence of (i) and (ii) follows in the framework of RG for a Kondo
model:\cite{poor}
Integrating out the p-h asymmetric part of the bath generates a potential scatterer.
This is seen most clearly in the case of a maximally asymmetric bath, $\rho(\w>0) = \rho_0 |\w/D|^r$ and $\rho(\w<0)=0$. Then the weak-coupling beta functions for the dimensionless running couplings
$j=\rho_0 J$ and $v=\rho_0 V$ read\cite{zawa_asy,florens07}
\begin{align}
\frac{d j}{d \ln \Lambda} = r j - \frac{j^2}{2} + 2 v j\,,~~~
\frac{d v}{d \ln \Lambda} = r v + \frac{3j^2}{16} + v^2
\end{align}
to second order, with $\Lambda$ being the running UV cutoff.
The $j^2$ term in the second equation is responsible for generating a
finite $v$ as soon as $j$ is non-zero.

These arguments do not depend on the value of the DOS exponent $r$, and so the different sources of p-h asymmetry remain equivalent for $r<0$. In this paper we thus employ a p-h symmetric bath, Eq.~(\ref{dos}), using the potential scattering $V\neq 0$ to tune p-h asymmetry.

We note, however, that the exchange coupling $J$ and potential scattering $V$ are not
independent parameters in the effective Kondo model \emph{if} derived from an underlying
Anderson model. As discussed in Ref.~\onlinecite{mf13}, the various phases of the Kondo
model in its most general form might not be accessible in the parent Andersonian system.

%%%%%%%%%%%%%%%%%%%%%%%%%%%%%%%%%%%%%%%%%%%%%%%%%%%%%

\subsection{Observables}
\label{sec:obs}

To pave the way for the discussion of the various phases of the power-law
Kondo model, we introduce relevant observables to characterize the impurity behavior.

%%%%%%%%%%%%%%%%%%%%%%%%%%%%%%%%%%%%%%%%%%%%%%%%%%%%%

\subsubsection{Susceptibilities}

Magnetic susceptibilities are obtained by coupling an external magnetic field $ \vec{H}_{\text{u}} $
to the bulk electronic degrees of freedom in $H_{b}$,
\begin{eqnarray}
&&- \vec{H}_{\text{u}} \cdot \sum_{{\vec k},\alpha\beta} c^\dagger_{{\vec k}\alpha} \vec{\sigma}_{\alpha\beta} c^{\phantom{\dagger}}_{{\vec k}\beta}
\end{eqnarray}
and coupling $\vec{H}_{\text{imp}}$ to the impurity spin in $H_{\text{K}}$ via,
\begin{eqnarray}
- \vec{H}_{\text{imp}} \cdot  \vec{S}\,.
\label{par2}
\end{eqnarray}
With these definitions, a spatially-uniform field applied
to the whole system corresponds to $\vec{H}_{\rm u} = \vec{H}_{\rm imp} \equiv \vec{H}$.
Response functions can be defined from second derivatives of the thermodynamic
potential, $\Omega = - T \ln Z$, in the standard way:\cite{impmag}
$\chi_{\rm{u},\rm{u}}$ measures the bulk response to a field applied
to the bulk, $\chi_{\rm{imp},\rm{imp}}$ is the impurity response to
a field applied to the impurity, and $\chi_{\rm{u},\rm{imp}}$
is the cross-response of the bulk to an impurity field.

The impurity contribution to the total susceptibility is defined as
\begin{eqnarray}
\label{chiimp}
\chi_{\rm imp}(T)
= \chi_{\rm imp,imp} + 2 \chi_{\rm u,imp} + (\chi_{\rm u,u} - \chi_{\rm u,u}^{\rm bulk})
\,,
\end{eqnarray}
where $\chi_{\rm u,u}^{\rm bulk}$ is the susceptibility of the bulk system in
absence of the impurity.
For a free unscreened impurity with spin $S=\frac{1}{2}$, one of course expects
$\chi_{\rm imp}(T\to 0) = S(S+1)/(3T)=1/(4T)$ in the low-temperature limit.
At critical points $\chi_{\rm imp}$ does {\em not} acquire an anomalous dimension \cite{ss}
(in contrast to $\chi_{\rm loc}$ below),
because it is a response function associated with the conserved total spin.
Thus one generically expects a Curie law susceptibility,
\begin{equation}
\lim_{T\to 0} \chi_{\rm imp}(T) = \frac{C_{\rm imp}}{T} \,,
\label{fract}
\end{equation}
but with an effective Curie constant $C_{\rm imp}$ that can in general take a non-trivial universal
value different from $S(S+1)/3$.
Apparently, Eq.~(\ref{fract}) can be interpreted as the Curie response of a
\emph{fractional} effective spin.\cite{impmag}
In the power-law Kondo model Eq.~\eqref{hk}, such a response is in fact realized in a stable
phase; another example of such a critical phase in a single-impurity model is in the
two-bath spin-boson (or XY-symmetric Bose-Kondo) model studied in
Refs.~\onlinecite{rg_bfk,guo12}.

The local impurity susceptibility is given by
\begin{equation}
\chi_{\rm loc}(T) = \chi_{\rm imp,imp}  \,.
\label{chiloc}
\end{equation}
In terms of the impurity spin autocorrelation function $\langle\langle \vec{S};\vec{S}
\rangle\rangle_{\w,T}\equiv
\chi_{\text{loc}}^{\prime}(\w,T)+i\chi_{\text{loc}}^{\prime\prime}(\w,T) $, the local
impurity susceptibility is obtained as the zero-frequency value, $\chi_{\rm
loc}(T)=\chi_{\text{loc}}^{\prime}(\w\rightarrow 0,T)$. At criticality, $\chi_{\rm
loc}$ typically follows a power law,
\begin{equation}
\lim_{T \rightarrow 0} \chi_{\text{loc}}(T) \propto T^{-1+\eta_\chi} \,.
\end{equation}
which defines the anomalous susceptibility exponent $\eta_\chi$.
For fixed points with hyperscaling properties --- this applies to all
intermediate-coupling fixed points of the power-law Kondo problem with $r<0$ --- the same
power-law behavior is realized as a function of frequency at zero-temperature, $\chi'_{\rm
loc}(\omega\rightarrow 0,T=0) \propto \omega^{-1+\eta_{\chi}}$, describing as such critical local-moment
fluctuations.

%%%%%%%%%%%%%%%%%%%%%%%%%%%%%%%%%%%%%%%%%%%%%%%%%%%%%

\subsubsection{Impurity entropy}

The impurity contribution to the entropy is defined as
\begin{equation}
\label{simp}
\s(T) = S - S^{\rm bulk}
\end{equation}
where $S^{\rm bulk}$ is the entropy of the system without the impurity. For most
phases of quantum impurity models, the residual impurity entropy, $S_{\rm imp}(T=0)$, is
a finite universal number of order unity, for example $\s=\ln 2$ for a free unscreened
impurity spin-$\tfrac{1}{2}$.

%%%%%%%%%%%%%%%%%%%%%%%%%%%%%%%%%%%%%%%%%%%%%%%%%%%%%

\subsection{NRG calculation of observables}

To obtain accurate numerical results for the Kondo model Eq.~\eqref{hk}, we employ Wilson's
NRG,\cite{nrg} generalized\cite{nrg_rev} to deal with the power-law conduction electron
DOS, Eq.~\eqref{dos}.
Impurity contributions to thermodynamic quantities, like $\tchi$ or $\s$, are obtained from
$\langle \hat{\Omega} \rangle_{\text{imp}} = \langle \hat{\Omega}
\rangle_{\text{tot}} - \langle \hat{\Omega} \rangle_{0}$, with
$\langle \hat{\Omega} \rangle_{\text{tot}}$ the thermal average of the
full impurity-coupled system, and $\langle \hat{\Omega} \rangle_{0}$
that of the free (`bath only') system. The full temperature dependence
is built up from information extracted from each iteration of the calculation.
Further details of the NRG algorithm can be found in Ref.~\onlinecite{nrg_rev}.
Throughout we use a discretization parameter $\Lambda=2$, and
$N_s=4000$ states are retained at each iteration.

%%%%%%%%%%%%%%%%%%%%%%%%%%%%%%%%%%%%%%%%%%%%%%%%%%%%%
%%%%%%%%%%%%%%%%%%%%%%%%%%%%%%%%%%%%%%%%%%%%%%%%%%%%%
%%%%%%%%%%%%%%%%%%%%%%%%%%%%%%%%%%%%%%%%%%%%%%%%%%%%%

\section{Phase diagram and RG flow}
\label{sec:phd}

We now summarize our results for the phase diagram and RG flow of the
power-law Kondo model with DOS exponent $-1<r<0$; qualitative aspects of these were
initially discussed in Ref.~\onlinecite{vb02}.

As for the pseudogap Kondo model with $r>0$, the fate of the impurity strongly depends
on the presence or absence of p-h symmetry. The qualitative behavior for
ferromagnetic coupling, $J<0$, moreover depends on the value of $r$: see Fig.~\ref{fig:flow}. Therefore, we
distinguish three intervals, bounded by $\bar{r}^\ast = -0.245 \pm 0.005$ and
$\bar{r}_{\rm max} = -0.264 \pm 0.001$, whose values have been obtained numerically by NRG (see also Sec.~\ref{critdim}).

%%%%%%%%%%%%%%%%%%%%%%%%%%%%%%%%%%%%%%%%%%%%%%%%%%%%%

\subsection{RG flow and fixed points}

\subsubsection{$J>0$}

\begin{figure}[!t]
\includegraphics[width=0.48\textwidth]{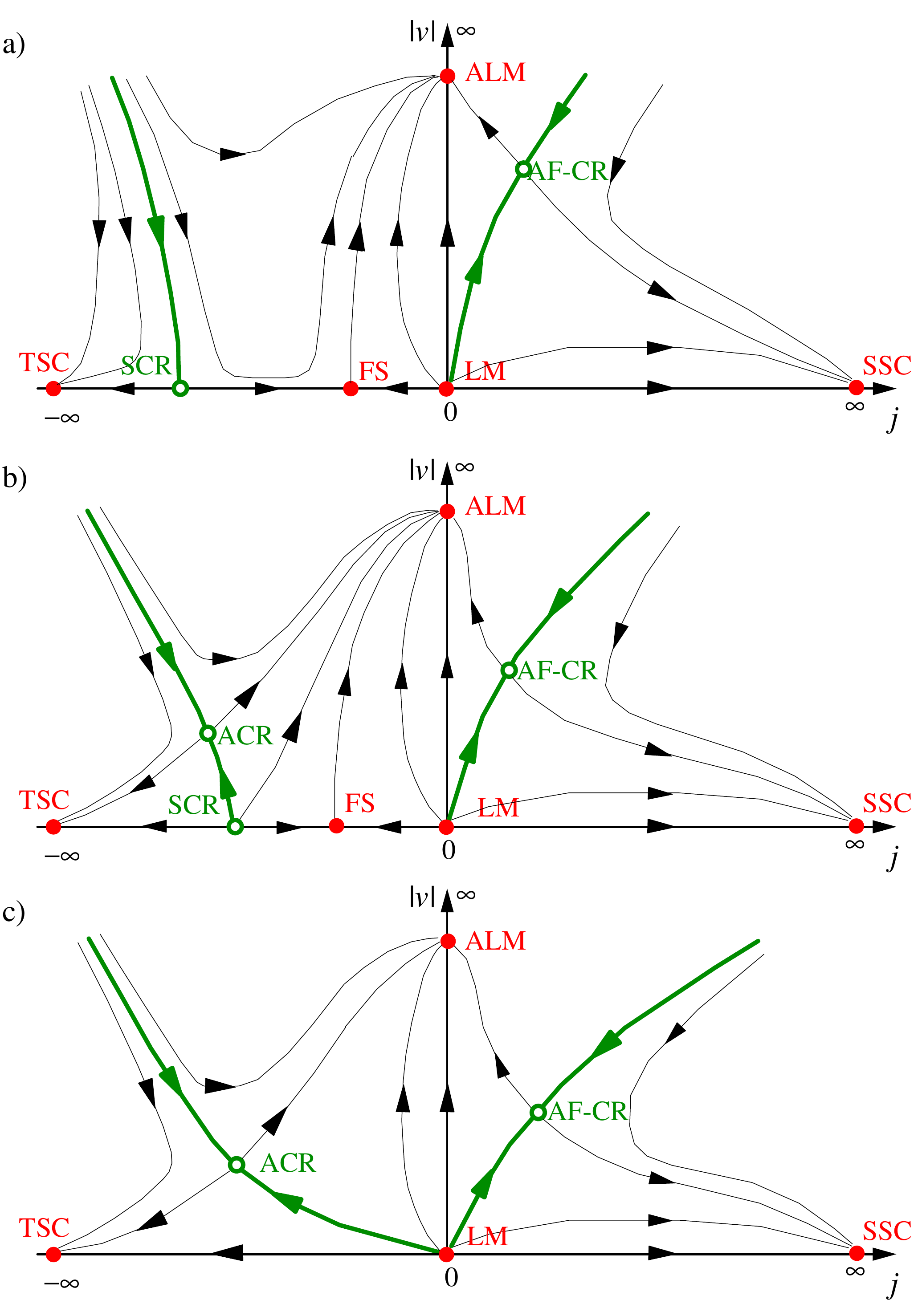}
\caption{
Schematic RG flows for the power-law Kondo model with $r<0$,
as deduced from NRG calculations. The horizontal axis denotes the (running) Kondo
coupling $j$ while the vertical axis denotes the (running) potential scattering $v$.
a) $\bar{r}^\ast<r<0$,
b) $\bar{r}_{\rm max}<r<\bar{r}^\ast$,
c) $-1<r<\bar{r}_{\rm max}$,
with NRG estimates of $\bar{r}^\ast = -0.245 \pm 0.005$, $\bar{r}_{\rm max} = -0.264 \pm
0.001$. The solid dots denote infrared stable fixed points, the open dots are critical
fixed points (labeled as in the text). FS denotes the infrared stable
intermediate-coupling fixed point with fractional spin.
} \label{fig:flow}
\end{figure}

\begin{figure}[t]
\includegraphics[width=0.42\textwidth]{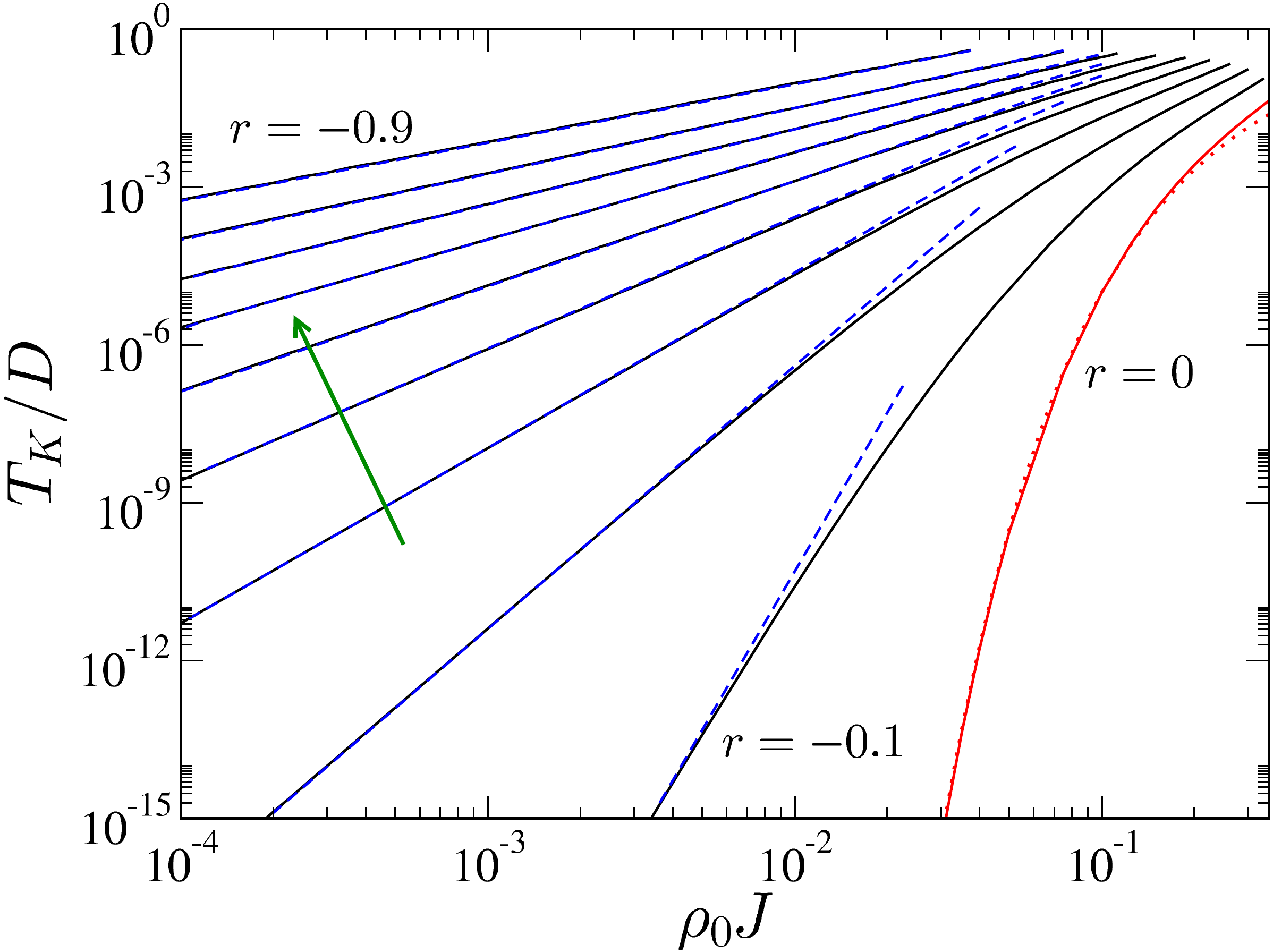}
\caption{
NRG results for the Kondo temperature, defined by $\s(\TK) = (\ln 2)/2$, as a function of the
Kondo coupling $J$ for a pure power-law bath DOS with different exponents $r=0,-0.1, -0.2, ..., -0.9$ decreasing in the direction of the arrow.
The dashed lines show the weak-coupling power-law of Eq.~\eqref{eq:tk}
[dotted line is the standard result, Eq.~\ref{eq:TKr0}, for the metallic case $r=0$].
}
\label{fig:tk}
\end{figure}

\begin{figure}[b]
\includegraphics[width=0.47\textwidth]{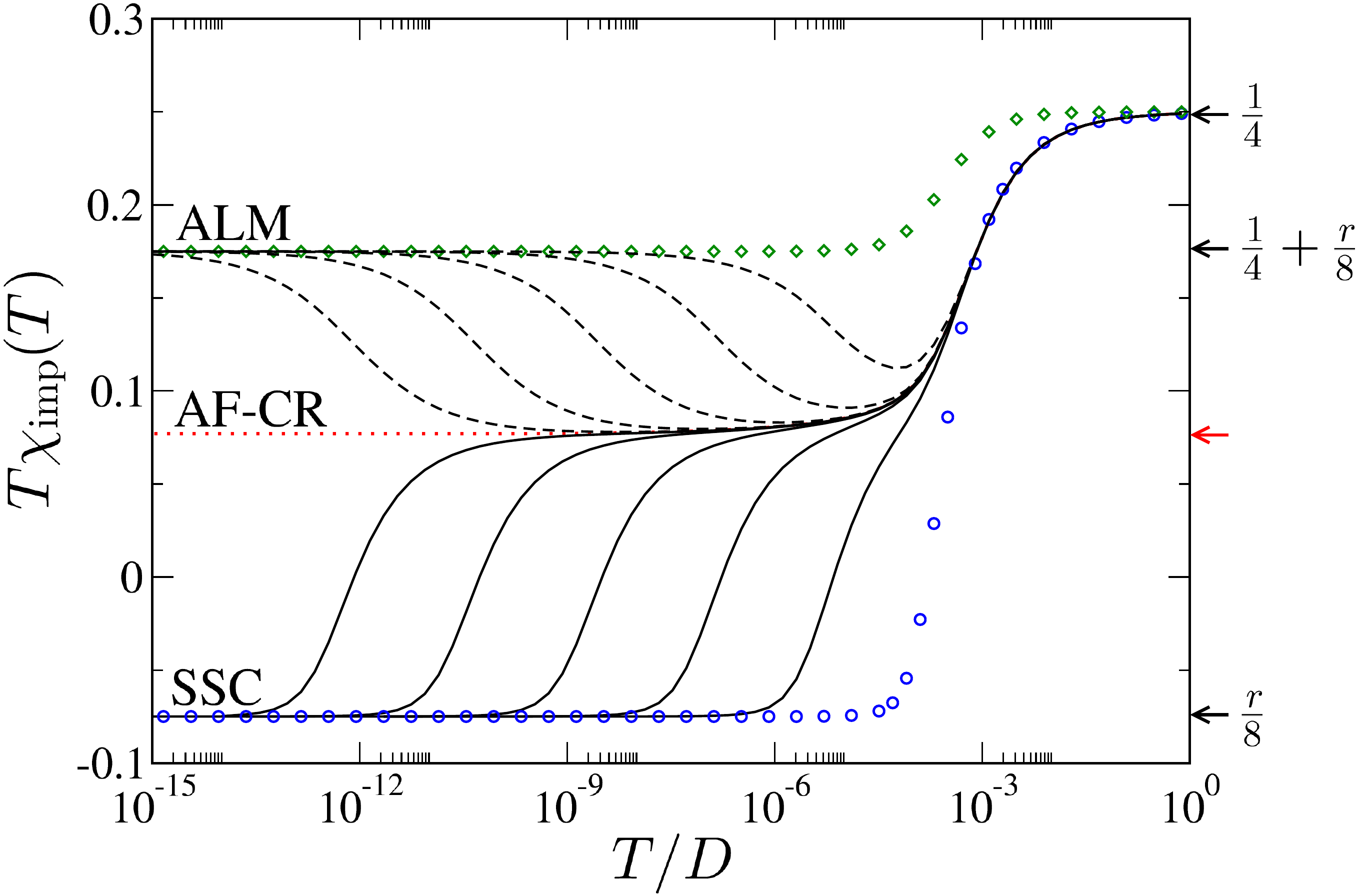}
\caption{
Impurity susceptibility $\tchi(T)$ vs. $T/D$ across the
QPT separating SSC and ALM phases, for antiferromagnetic Kondo
coupling $J=0.01$ and $r=-0.6$. The AF-CR critical point (red dotted line) is attained
at $V_c\simeq 0.01025...$. The transition is approached
progressively from the SSC phase (solid lines, $V<V_c$) or the ALM phase (dashed lines,
$V>V_c$) with $V=V_c \pm \delta V$ and $\delta V=10^{-3}, 10^{-4}, 10^{-5}, 10^{-6},
10^{-7}$. For comparison, circles are for the direct crossover with $J=0.01$ and
$V=0$, while diamonds are for $J=0$ but $V=0.01$. Arrows indicate the fixed point
values.
}
\label{fig:afqpt}
\end{figure}

For antiferromagnetic Kondo coupling, we observe screening in the p-h symmetric case
--- controlled by the singlet strong-coupling (SSC) fixed point --- for any value of $J>0$. Kondo screening is of course well-known to arise already in the standard metallic case, $r=0$; and so screening is also naturally expected in the power-law case. The enhanced DOS near the Fermi level should give rise to a larger Kondo temperature, which can itself be identified by integrating the weak coupling RG flow (see Eq.~\eqref{eq:wcrgflow} of Sec.~\ref{sec:critfp}). The Kondo strong coupling behavior is associated with a divergence of the renormalized coupling, which arises on the scale,
\begin{eqnarray}\label{eq:fullTKrg}
\TK \sim D \left(1-\frac{r}{\rho_0 J} \right)^{1/r}\;.
\end{eqnarray}
In the limit $r\rightarrow 0$, one thus recovers the standard result,\cite{hewson}
\begin{eqnarray}\label{eq:TKr0}
\TK\overset{r \rightarrow 0}{\sim} D\exp(-1/\rho_0 J)\;,
\end{eqnarray}
while the diverging power-law DOS with $-1<r<0$ yields instead an \emph{enhanced} power-law dependence of the Kondo scale,
\begin{equation}
\label{eq:tk}
\TK \overset{\rho J \ll |r|}{\sim} D \left (\frac{\rho_0 J}{-r} \right)^\alpha
\end{equation}
with leading-order exponent $\alpha = -1/r$. For larger values of $\rho_0 J$, one observes deviations consistent with Eq.~\eqref{eq:fullTKrg}.
These results are confirmed numerically in Fig.~\ref{fig:tk}.

For finite p-h asymmetry, the Kondo-screened state competes with a maximally p-h asymmetric
unscreened state, controlled by the asymmetric local-moment (ALM) fixed point.
As a consequence, no screening occurs for small $J$ and large $V$ --- see Fig.~\ref{fig:flow}.
At ALM, the impurity is asymptotically decoupled from the bath, and the conduction electron site localized at the impurity position is either doubly occupied or empty (depending on the sign of the bare $V$). The transition between SSC and ALM is controlled by an antiferromagnetic critical (AF-CR)
fixed point, as shown in Fig.~\ref{fig:afqpt}.

\begin{figure}[t]
\includegraphics[width=0.47\textwidth]{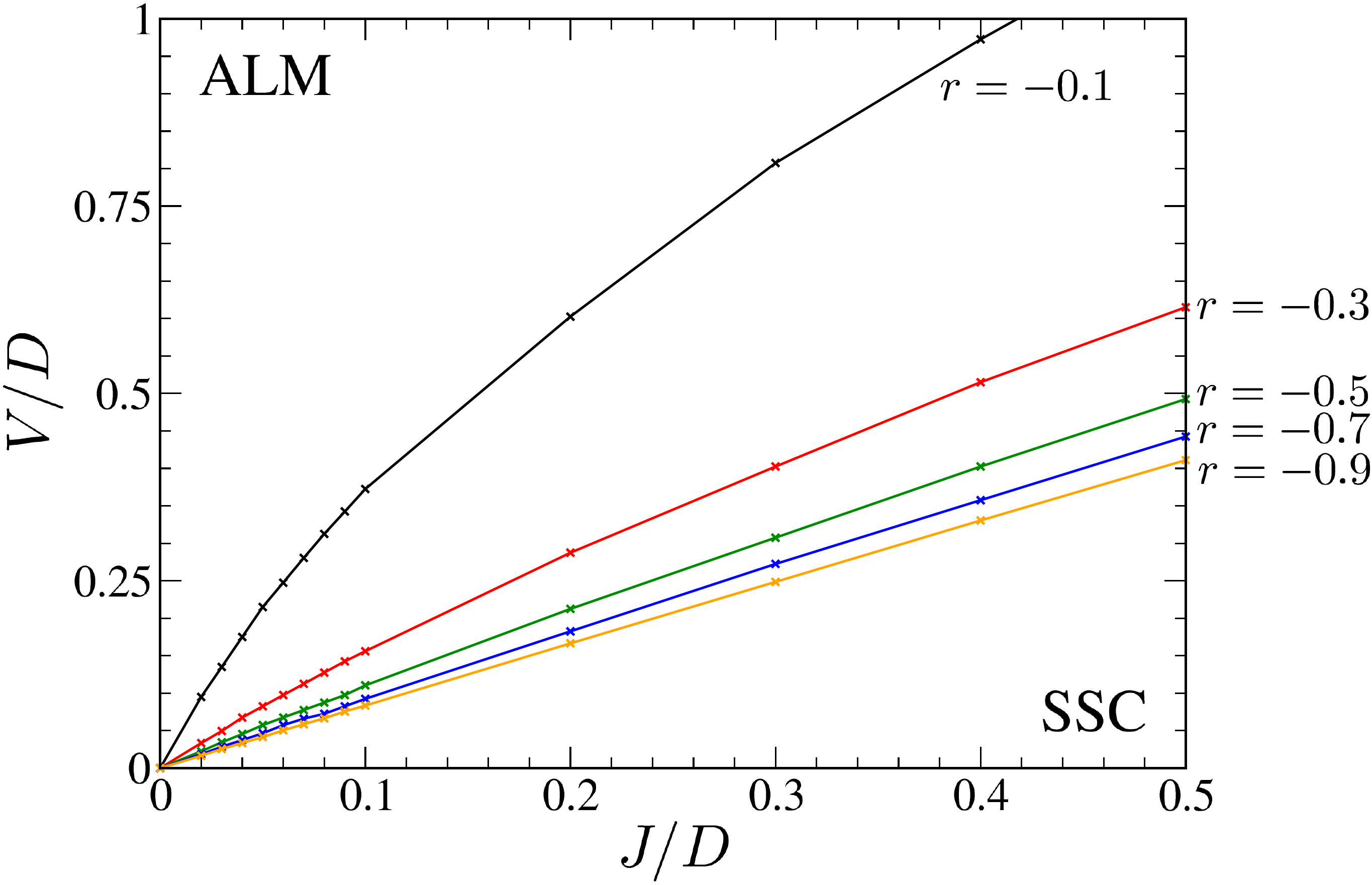}
\caption{
Exact phase boundary between SSC and ALM phases of the Kondo model with
pure power-law DOS Eq.~\eqref{dos} and
antiferromagnetic Kondo coupling, obtained using NRG.
}
\label{fig:phaseboundary}
\end{figure}

A quantitative ground-state phase diagram obtained using NRG is presented in
Fig.~\ref{fig:phaseboundary}, showing the phase boundary between SSC and ALM for
different values of $r<0$.

\begin{figure}[t]
\includegraphics[width=0.4\textwidth]{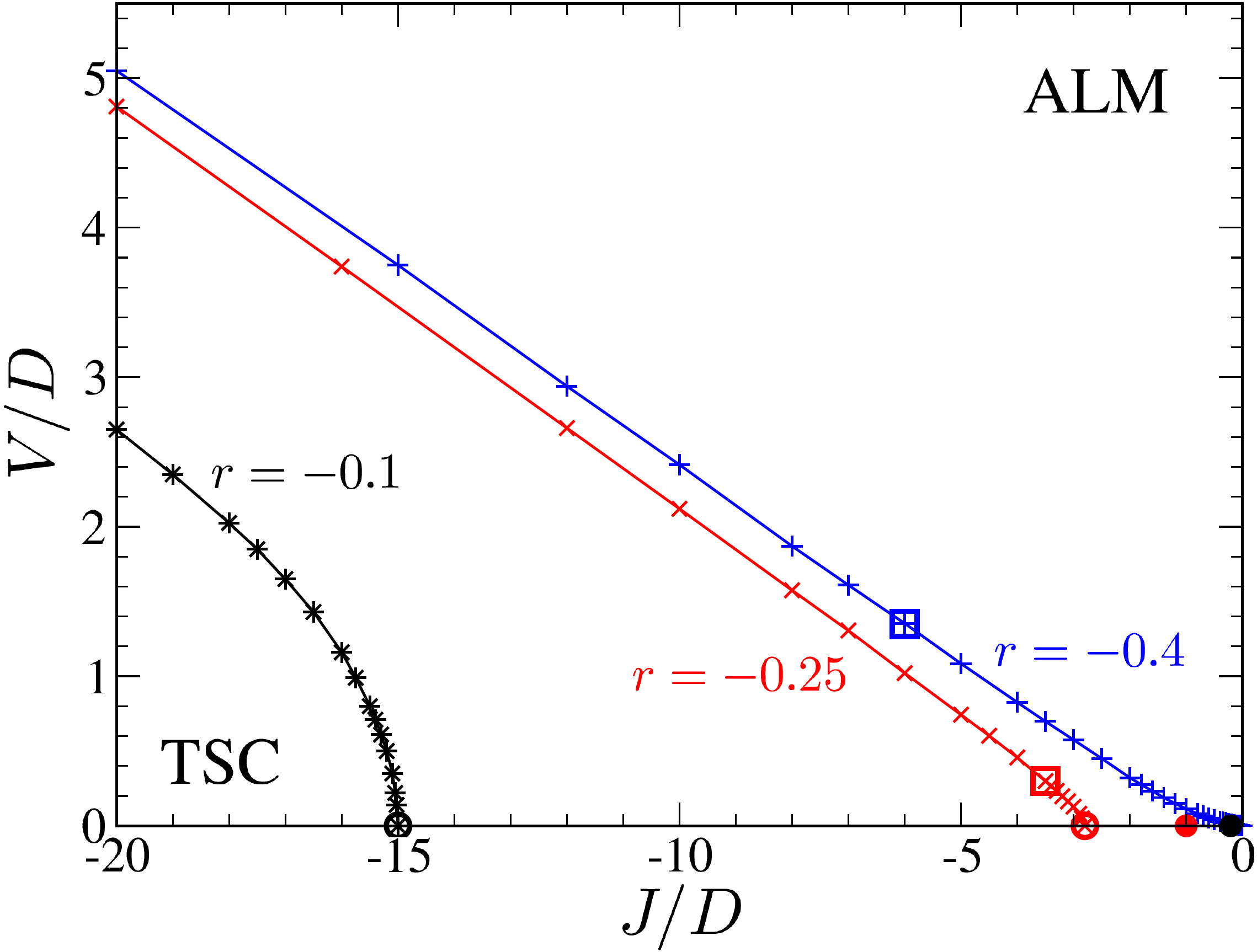}
\caption{
Exact phase boundaries between TSC and ALM phases of the Kondo model
with pure power-law DOS Eq. (2) and ferromagnetic Kondo coupling,
obtained using NRG.
Shown for bath exponents $r=-0.1$, $-0.25$ and $-0.4$ (black stars,
red crosses, and blue plusses, respectively), characteristic of the
three distinct regimes. Open circles represent the SCR fixed point [at
$(J,V)\simeq (-15.0,0)$ and $(-2.8,0)$ for $r=-0.1$ and $-0.25$], filled
circles correspond to the FS fixed point [at $(J,V)\simeq (-0.2,0)$ and
$(-1,0)$ for $r=-0.1$ and $-0.25$], and squares are for the ACR fixed
point [at $(J,V)\simeq (-3.5,0.3)$ and $(-6,1.35)$ for $r=-0.25$ and
$-0.4$].}
\label{fig:phaseboundaryfm}
\end{figure}

\subsubsection{$J<0$, $\bar{r}^\ast<r<0$}

In the p-h symmetric case, small ferromagnetic Kondo coupling flows to a finite
intermediate value, realizing the fractional-spin (FS) phase advertised in
Ref.~\onlinecite{vb02}. However, for larger $(-J)$, one obtains a stable phase where the
impurity spin-$\tfrac{1}{2}$ and the bath site to which it is coupled lock together in a
\emph{triplet} configuration. This state is described by the triplet strong-coupling
(TSC) fixed point. FS and TSC are themselves separated by a QPT controlled by a p-h
symmetric critical (SCR) fixed point.

With p-h asymmetry, FS becomes unstable, generating a flow to ALM. However, SCR is stable
for $\bar{r}^\ast<r<0$: As a result, the transition between TSC and ALM, accessed on
tuning potential scattering $V$ for large ferromagnetic $J$, is also controlled by SCR,
which is therefore multicritical in nature --- see Fig.~\ref{fig:flow}(a) for the
qualitative RG flow, and star points of Fig.~\ref{fig:phaseboundaryfm} for the exact
phase boundary obtained by NRG.

\subsubsection{$J<0$, $\bar{r}_{\rm max}<r<\bar{r}^\ast$}

Upon decreasing $r$ below $\bar{r}^\ast$, a new p-h asymmetric critical (ACR) fixed point
splits off from SCR. The transition between TSC and ALM, accessed on tuning $V$, is now
controlled by ACR. However, in the p-h symmetric case $V=0$, the transition between TSC
and FS, accessed on tuning $J$, remains controlled by SCR. See  Fig.~\ref{fig:flow}(b)
and cross points in Fig.~\ref{fig:phaseboundaryfm}.

\subsubsection{$J<0$, $-1<r<\bar{r}_{\rm max}$}

\begin{figure}[t]
\includegraphics[width=0.47\textwidth]{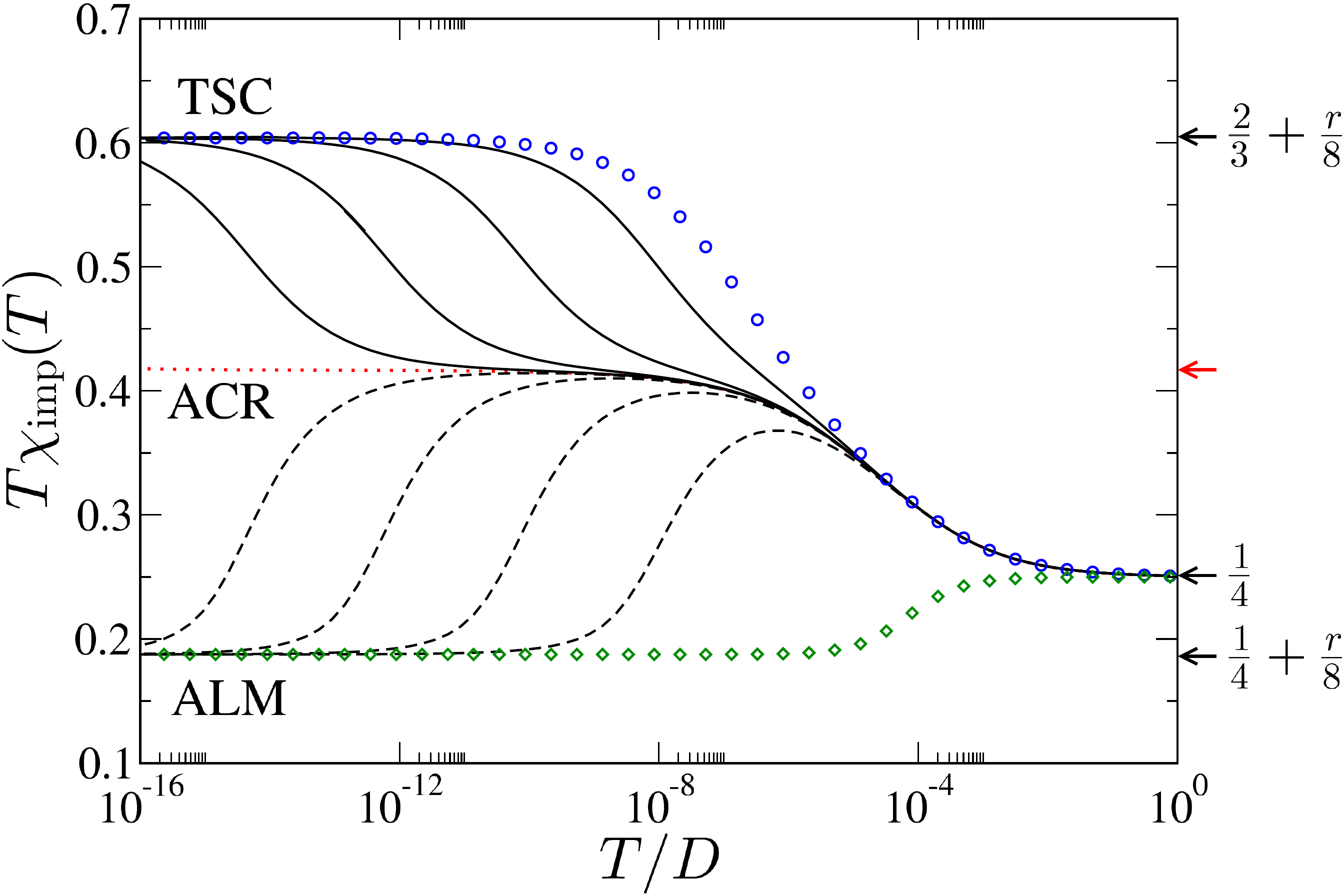}
\caption{
$\tchi(T)$ vs. $T/D$ as in Fig.~\ref{fig:afqpt}, but now
across the QPT separating TSC and ALM phases, for ferromagnetic
Kondo coupling $J=-0.01$ and $r=-0.5$. The ACR critical point (red dotted line) is
attained at $V_c\simeq 0.001106...$. The transition is
approached progressively from the TSC phase (solid lines, $V<V_c$) or the ALM phase
(dashed lines, $V>V_c$) with $V=V_c \pm \delta V$ and $\delta V=10^{-4}, 10^{-5}, 10^{-6},
10^{-7}$. For comparison, circles are for the direct crossover with $J=-0.01$ and
$V=0$, while diamonds are for $J=0$ but $V=0.01$.
}
\label{fig:fmqpt}
\end{figure}

For $r\to\bar{r}_{\rm max}^+$, the p-h symmetric fixed points FS and SCR merge and
annihilate, such that for $r<\bar{r}_{\rm max}$ \emph{any} ferromagnetic Kondo coupling
flows to large values and produces the triplet state described by TSC. See
Fig.~\ref{fig:flow}(c) and  Fig.~\ref{fig:phaseboundaryfm}. The transition between TSC
and ALM remains controlled by ACR, as demonstrated by the evolution of thermodynamics
near criticality in Fig.~\ref{fig:fmqpt}.

%%%%%%%%%%%%%%%%%%%%%%%%%%%%%%%%%%%%%%%%%%%%%%%%%%%%%

\subsection{Mappings to positive-$r$ models}

In the course of the paper we establish and exploit various mappings between (a certain
parameter regime of) the power-law Kondo model with negative $r$ and variants of the
pseudogap Kondo model with positive DOS exponent $r'=-r$. A comprehensive analytical
understanding of these pseudogap Kondo models has been achieved in
Refs.~\onlinecite{vf04,fv04,kv04,serge05,DELlma}, and we will make use of the corresponding results.

As previous papers\cite{GBI,vf04,fv04,serge05,DELlma} label positive-$r$
phases and fixed points using acronyms similar to the ones employed here for the negative-$r$ case,
in the following we distinguish them by using primed labels for fixed points of the positive-$r$ effective models, e.g., LM$'$, SSC$'$ etc.

%%%%%%%%%%%%%%%%%%%%%%%%%%%%%%%%%%%%%%%%%%%%%%%%%%%%%
%%%%%%%%%%%%%%%%%%%%%%%%%%%%%%%%%%%%%%%%%%%%%%%%%%%%%
%%%%%%%%%%%%%%%%%%%%%%%%%%%%%%%%%%%%%%%%%%%%%%%%%%%%%

\section{Trivial fixed points}
\label{sec:trivfp}

As can be seen from Fig.~\ref{fig:flow}, the power-law Kondo problem displays a number of
stable phases, controlled by fixed points where the renormalized couplings $|j|$ and $|v|$ have flowed to either zero or infinity. These fixed points can be fully understood analytically, as
explained below, with important thermodynamic properties summarized in
Table~\ref{tab:trivfp}.

\begin{table}[b]
\begin{center}
\begin{tabular}{|c|c|c|}
\hline
Fixed point & Entropy, $\s$ & Curie constant, $C_{\text{imp}}$ \\
\hline \hline
\multirow{2}{*}{Local Moment}  & \multirow{2}{*}{$\ln 2$} & \multirow{2}{*}{$\frac{1}{4}$}\\
(LM) & & \\
\hline
\multirow{2}{*}{Symmetric Strong}  & \multirow{2}{*}{$2r \ln 2$} & \multirow{2}{*}{$\frac{r}{8}$}\\
 Coupling (SSC) & & \\
\hline
\multirow{2}{*}{Asymmetric Local}   & \multirow{2}{*}{$\ln 2 + 2r \ln 2$} & \multirow{2}{*}{$\frac{1}{4} + \frac{r}{8}$}\\
 Moment (ALM) & & \\
\hline
\multirow{2}{*}{Triplet Strong} & \multirow{2}{*}{$\ln 3 + 2r \ln 2$} & \multirow{2}{*}{$\frac{2}{3} + \frac{r}{8}$}\\
Coupling (TSC)  & & \\
\hline
\end{tabular}
\end{center}
\caption{
Exact thermodynamic properties of the trivial fixed points of the power-law Kondo model.
}
\label{tab:trivfp}
\end{table}

%%%%%%%%%%%%%%%%%%%%%%%%%%%%%%%%%%%%%%%%%%%%%%%%%%%%%

\subsection{Resonant-level models and chain representation}
\label{sec:chain}

An effective model which plays a central role in the following analysis is the
non-interacting resonant-level model (RLM), described by the Hamiltonian
$H_{\text{RLM}}=H_b+H_l$ with
\begin{equation}
\label{hrlm}
H_l = V_0 \sum_{\sigma}(f_\sigma^\dagger c^{\phantom{\dagger}}_{0\sigma} + \text{H.c.})
\end{equation}
where $H_b$ is the p-h symmetric conduction electron Hamiltonian, with power-law DOS given by Eq.~\eqref{dos}; and $f_\sigma$ is an operator for the additional resonant level.
This model is of course exactly solvable and has been analyzed before in
Refs.~\onlinecite{GBI,fv04,DELpt}. For $r<1$ it displays remarkable thermodynamic
properties, which can be calculated directly from the $f$-electron propagator
in the low-energy limit,
\begin{equation}
\label{gfrlm}
G_f(i\omega_n)^{-1} = i\omega_n - i A_0\,{\rm
sgn}(\omega_n)\,|\w_n|^r - A_1 i\w_n
\end{equation}
where $A_0 = (\pi \rho_0 V_0^2)/[D^r \cos(\pi r/2)]$ and $A_1= (2\rho_0 V_0^2) / [D(1-r)]$.
The low-energy behavior can also be understood more physically in terms of the
RG flow to an intermediate-coupling fixed point.\cite{fv04}
In particular, the impurity entropy associated with the resonant level is $\s^{\rm RLM} =
2 r \ln 2$ and the susceptibility is $\tchi^{\rm RLM} = r/8$ --- these equations were
originally derived\cite{GBI,fv04,DELpt} for $0\leq r <1$, but in fact continue
to hold for $-1<r<0$.

Interestingly, both the entropy and Curie constant are {\em negative} for the $r<0$
situation considered here; this reflects the fact that coupling the resonant-level
impurity to the bath with divergent DOS removes low-energy states from the system. We
stress that since $\s$ is defined as the difference in entropy of two
thermodynamically-large systems, a negative $\s$ does not contradict any fundamental laws
of thermodynamics.

\begin{figure}[!t]
\includegraphics[width=0.4\textwidth]{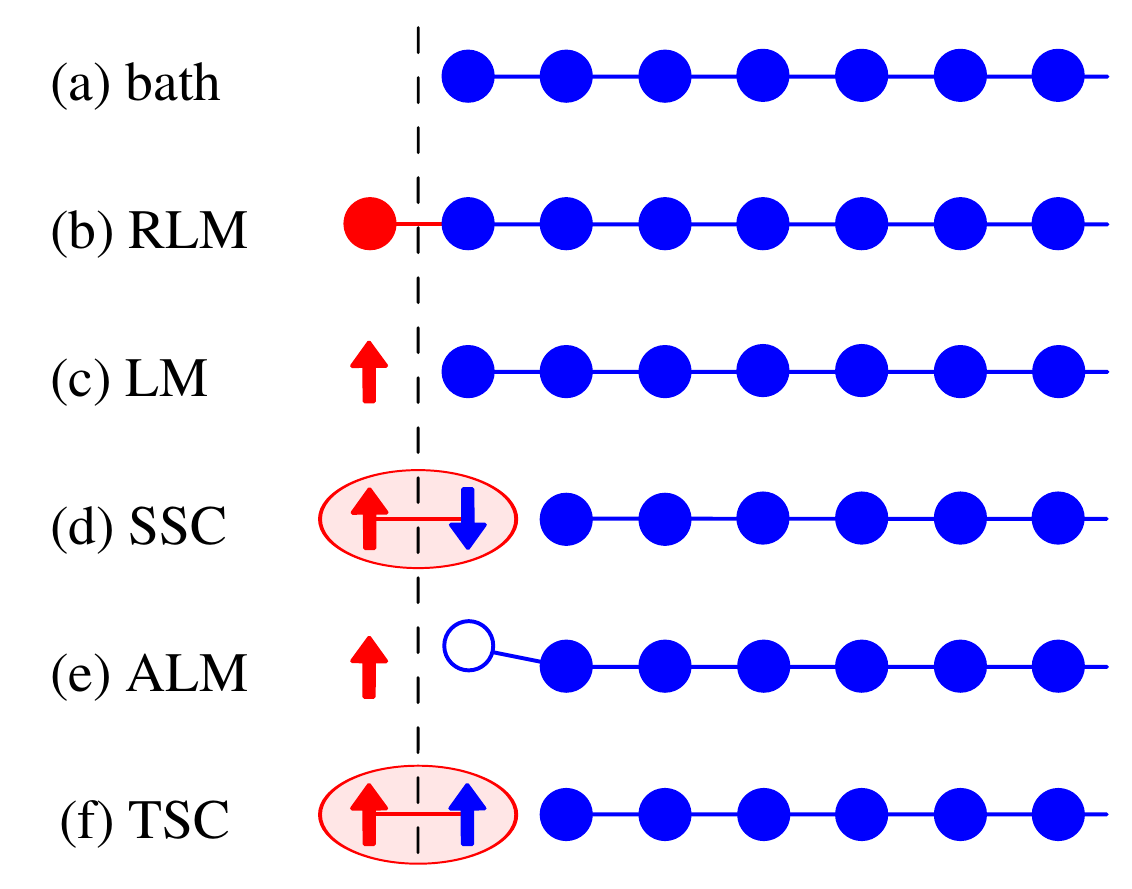}
\caption{
Chain representation of the trivial fixed points of the power-law Kondo model.
Circles denote fermion sites; the dashed line separates the bath sites (right) from the
impurity (left). Full (open) circles denote sites with (without) p-h symmetry; the
ellipses indicate strong-coupling singlet or triplet states; for details see text.
}
\label{fig:chain}
\end{figure}

\begin{figure*}[!bt]
\includegraphics[width=0.85\textwidth]{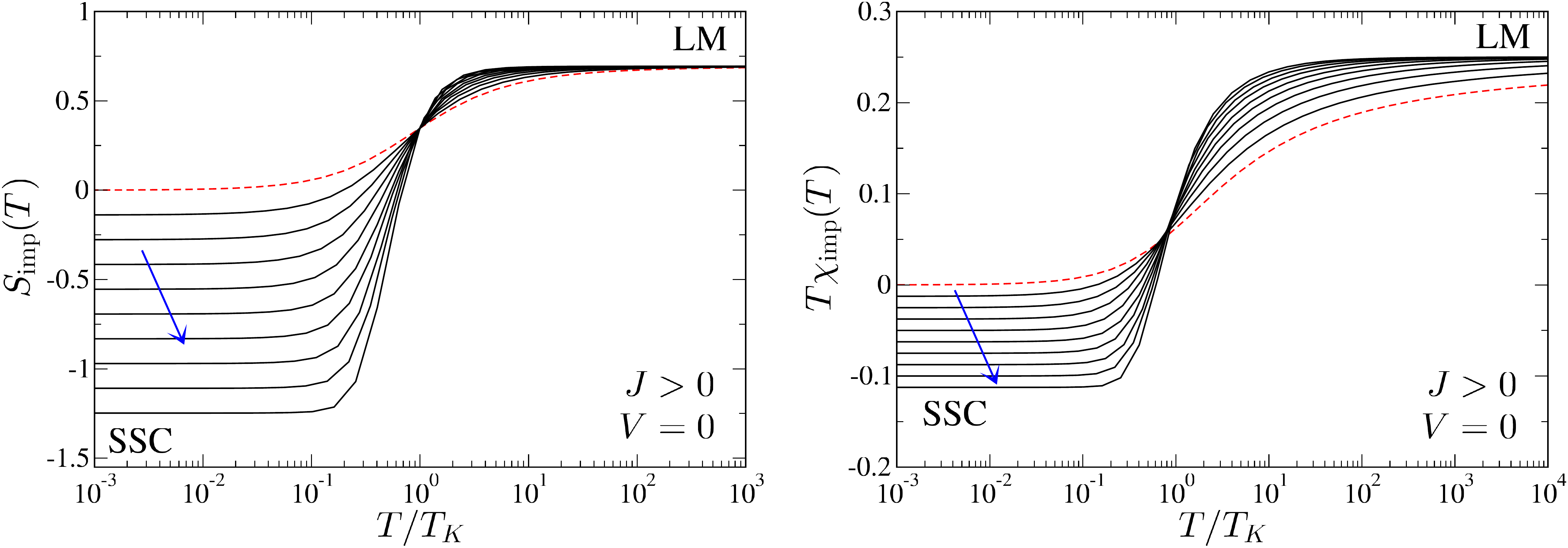}
\caption{
Universal scaling curves for impurity entropy, $\s(T)$, and impurity susceptibility, $\tchi(T)$,
in the case of antiferromagnetic Kondo coupling
$J>0$ (but $V=0$) for various bath exponents $r=0, -0.1, -0.2, ...  , -0.9$ (following the arrow).
The temperature is plotted as $T/\TK$, with the Kondo temperature $\TK$ defined via $\s(\TK) = (\ln
2)/2$.
In the limit $T\rightarrow 0$, we have $\s = 2r\ln 2$ and $\tchi = r/8$
at the SSC fixed point.
}
\label{fig:af_ent}
\end{figure*}

An intuitive understanding of various fixed points can be achieved using a representation
of the bath in terms of a semi-infinite p-h symmetric tight-binding chain,
Fig.~\ref{fig:chain}(a). The first site of the chain is precisely the bath site localized at the impurity, ${\vec r}_0 = 0$ (the $c_{0\sigma}$ orbital of Eq.~\eqref{hk}). Starting from this site, the rest of the bath is tridiagonalized to yield the 1d chain representation (similarly, a discretized version of the model yields the ``Wilson chain'' representation exploited in NRG\cite{nrg,nrg_rev}).

The resonant-level model, Eq.~\eqref{hrlm}, then simply corresponds to a chain with one
additional site, Fig.~\ref{fig:chain}(b). From Eq.~\eqref{gfrlm} it follows that the local DOS of the resonant level (now the leading site of the extended chain) has a form $\propto |\w|^{-r}$ at low energies. Indeed, by obvious extension, adding additional sites leads to an alternation of the low-energy behavior, such that a chain with an even (odd) number $n$ of extra sites displays a $|\w|^{r}$ ($|\w|^{-r}$) low-energy DOS. This argument similarly holds for negative $n$ (removed sites). The implication for thermodynamic quantities is that as $T\rightarrow 0$, removing one site from the bath yields the same excess entropy, $2r\ln 2$, as adding one site. This can be understood as follows: a shortened chain with $n=-1$ has local DOS of the terminal site with $|\w|^{-r}$ behavior. Adding an extra site (restoring the original chain, $n=0$) gives an additional entropy of $(-2r\ln 2)$. The entropy of the full chain is larger by $(-2r\ln 2)$ as compared with the chain with a site removed.

The chain representation is exploited below to study the various trivial fixed points of the problem.

%%%%%%%%%%%%%%%%%%%%%%%%%%%%%%%%%%%%%%%%%%%%%%%%%%%%%

\subsection{LM}
\label{sec:lm}

The p-h symmetric local-moment fixed point at $j=v=0$ corresponds to a decoupled spin-$\tfrac{1}{2}$
impurity, Fig.~\ref{fig:chain}(c). Consequently, we have $\s^{\rm LM}=\ln 2$ and
$\tchi^{\rm LM}=1/4$.

%%%%%%%%%%%%%%%%%%%%%%%%%%%%%%%%%%%%%%%%%%%%%%%%%%%%%

\subsection{SSC}
\label{sec:ssc}

The p-h symmetric singlet strong-coupling fixed point describes a Kondo-screened impurity, with $j=+\infty$. In the chain representation, this corresponds to a local singlet (formed between the impurity spin and the singly-occupied local bath site to which it is coupled), disconnected from a free chain with one site removed, Fig.~\ref{fig:chain}(d).
As argued above, a chain with one site removed is thermodynamically equivalent to a chain
with one site added.
Since the singlet is inert, this establishes the equivalence of the SSC fixed
point with the resonant-level model (of course, this equivalence is also well-known in the context of the standard metallic Kondo problem\cite{hewson}).
As a result, we have $\s^{\rm SSC}=2r\ln 2$ and $\tchi^{\rm SSC}=r/8$, as derived
before\cite{GBI,fv04,DELpt} for positive $r$.

Fig.~\ref{fig:af_ent} shows numerical results, obtained using NRG, for both $\s$ and
$\tchi$ for the finite-temperature crossover from LM to SSC, which correspond to the RG
flow along the p-h symmetric positive $j$ axis in Fig.~\ref{fig:flow}. The $T\to0$
values of $\s$ and $\tchi$ are in perfect agreement with the analytical predictions for
SSC.

%%%%%%%%%%%%%%%%%%%%%%%%%%%%%%%%%%%%%%%%%%%%%%%%%%%%%

\subsection{ALM}
\label{sec:alm}

The p-h asymmetric local-moment fixed point corresponds to a decoupled spin-$\tfrac{1}{2}$
impurity, $j=0$, supplemented by a divergent potential scattering term, $|v|=\infty$.
The latter effectively eliminates the site $\vec{r}=0$ from the bath,
Fig.~\ref{fig:chain}(e). Consequently, the thermodynamic properties are that of a free
spin plus a resonant level model,
$\s^{\rm ALM}=(1+2r)\ln 2$ and $\tchi^{\rm ALM}=(2+r)/8$.

When potential scattering of strength $V$ is applied to the last site of the chain (${\vec r}_0 = 0$), the problem is exactly solved using the $T$ matrix formalism.\cite{hewson} We have
\begin{eqnarray}
G(\vec{r}=0,\omega_n)=G^0 + G^0 V (1- G^0 V)^{-1} G^0
\end{eqnarray}
where $G^0 = G^0(\vec{r}=0,\omega_n)$ is the local $\vec{r}=0$ Green function
without potential scattering.
Using the DOS in Eq.~\eqref{dos} with negative $r$ yields,
\begin{eqnarray}
\label{tdos}
G(\vec{r}=0,\omega_n)= - \frac{1}{V} - i B_0 {\rm sgn}(\w_n) |\w_n|^{-r} + \mathcal{O}(\w_n^{-2r})
\end{eqnarray}
with $B_0 = D^r \cos(\pi r/2) / (V^2 \rho_0 \pi)$.
Thus $V\neq 0$ has the singular effect of converting the bath DOS from $|\w|^r$ to
$|\w|^{-r}$ at low energies. This results in an instability of LM toward finite potential scattering, and thereby the flow to $|v|=\infty$ corresponding to ALM, Fig.~\ref{fig:flow}.

%%%%%%%%%%%%%%%%%%%%%%%%%%%%%%%%%%%%%%%%%%%%%%%%%%%%%

\subsection{TSC}
The p-h symmetric triplet strong-coupling fixed point describes a spin-1 object,
formed due to strong ferromagnetic coupling $j=-\infty$ between the spin-$\tfrac{1}{2}$ impurity and the singly-occupied bath site to which it is coupled. At the TSC fixed point, this local triplet is disconnected from a chain with one site removed, Fig.~\ref{fig:chain}(f).
Therefore, one naturally gets contributions from both the spin-1 and the resonant level,
$\s^{\rm TSC}=\ln 3+2r\ln 2$ and $\tchi^{\rm TSC}=2/3 + r/8$.

\begin{table*}[!t]
\begin{center}
\begin{tabular}{|c|c|c|c|}
\hline
\multicolumn{2}{|c|}{Fixed point} & Entropy, $\s$ & Curie constant, $C_{\text{imp}}$ \\
\hline \hline
\multicolumn{2}{|c|}{Fractional Spin (FS)}  & $\ln 2 + \mathcal{O}(r^3)$ & $\frac{1-r}{4} + \mathcal{O}(r^2)$\\[1ex]
\hline
\multicolumn{2}{|c|}{Asymmetric-Critical (ACR)} & $\ln 5 +2r\ln 2- \frac{24\ln 2}{25}(1+r) + \mathcal{O}((1+r)^2)$ & $\frac{4+r}{8} - 0.03227 (1+r) + \mathcal{O}((1+r)^2)$ \\[1ex]
\hline
\multicolumn{2}{|c|}{Symmetric-Critical (SCR)}  & $\ln 3 + 2r \ln 2 + \mathcal{O}(r^3)$ & $\frac{16+19r}{24} + \mathcal{O}(r^2)$\\[1ex]
\hline
\multirow{2}{*}{Antiferromagnetic-Critical}  & $r^*<r<0$ &$\ln 2 + 2r \ln 2 + \mathcal{O}(r^3)$ & $\frac{2+3r}{8} + \mathcal{O}(r^2)$ \\[1ex]
\cline{2-4}
 (AF-CR) & $-1<r<r^*$ & $\ln 3 +2r\ln 2- \frac{8\ln 2}{9}(1+r) + \mathcal{O}((1+r)^2)$ & $\frac{4+3r}{24} - 0.02988 (1+r) + \mathcal{O}((1+r)^2)$ \\[1ex]
\hline
\end{tabular}
\end{center}
\caption{
Thermodynamic properties of the intermediate-coupling fixed points of the power-law Kondo
model, obtained by perturbative RG methods.
}
\label{tab:critfp}
\end{table*}

%%%%%%%%%%%%%%%%%%%%%%%%%%%%%%%%%%%%%%%%%%%%%%%%%%%%%
%%%%%%%%%%%%%%%%%%%%%%%%%%%%%%%%%%%%%%%%%%%%%%%%%%%%%
%%%%%%%%%%%%%%%%%%%%%%%%%%%%%%%%%%%%%%%%%%%%%%%%%%%%%

\section{Intermediate-coupling fixed points and RG}
\label{sec:critfp}

We now turn our attention to the intermediate-coupling fixed points of the power-law
Kondo model. These include the critical fixed points SCR, ACR, and AF-CR as well as the
stable fixed point FS, controlling the fractional spin phase.

In the following, we employ RG techniques to calculate perturbatively critical
properties. Specifically, we exploit a double expansion in a coupling constant and the deviation from a critical ``dimension'', which here corresponds to a special value of the bath exponent
$r$. A summary of thermodynamic results is presented in Table~\ref{tab:critfp}.

%%%%%%%%%%%%%%%%%%%%%%%%%%%%%%%%%%%%%%%%%%%%%%%%%%%%%

\subsection{FS: Kondo expansion in $J$}
\label{sec:fs}

The FS fixed point, which exists for $\bar{r}_{\rm max}<r<0$, can be accessed in an expansion
in the Kondo coupling $J$ around the LM fixed point. This is equivalent to Anderson's
poor man's scaling \cite{poor} adapted to the power-law Kondo model.\cite{withoff}
Expressed in $\beta$ functions for the dimensionless running couplings $j=\rho_0 J$ and
$v=\rho_0 V$ we have
\begin{eqnarray}\label{eq:wcrgflow}
\frac{d j}{d \ln \Lambda} = j (r - j) + {\cal O}(j^3) ~,~~~~
\frac{d v}{d \ln \Lambda} = r v .
\label{rgeq}
\end{eqnarray}
For $r<0$, small $|j|$ grows for {\em both} antiferromagnetic $j>0$ and ferromagnetic $j<0$ Kondo coupling. In particular, the RG equations predict (to second order) an infrared stable
fixed point at $v=0$ and $j^\ast = - |r|$ on the ferromagnetic side --- this is the FS fixed point.

Its properties are perturbatively accessible in a double expansion in $j$ and $r$ (around $j=0$ and $r=0$). The calculation parallels that for the critical fixed point in the $r>0$ case,\cite{fv04,kv04} and so we only quote the results here. The residual entropy is,
\begin{equation}
\label{sfs}
\s^{\rm FS} = \ln 2 + \mathcal{O}(r^3);
\end{equation}
the perturbative correction $\propto r^3$ was calculated in Ref.~\onlinecite{kv04},
but turned out to be unobservable numerically in the positive-$r$ case. The impurity
susceptibility follows
\begin{equation}
\label{tchifs}
\tchi^{\rm FS} = \frac{1 - r}{4} + \mathcal{O}(r^2).
\end{equation}
The predictions Eqs.~\eqref{sfs} and \eqref{tchifs} are compared to NRG data in
Figs.~\ref{fig:FS}(a) and (b) as the dashed lines, and agree essentially perfectly for small $r$.
The anomalous exponent of the local susceptibility evaluates to $\eta^{\rm FS}_\chi =
r^2 + \mathcal{O}(r^3)$. FS is unstable with respect to finite p-h asymmetry, with the scaling
dimension being ${\rm dim}[v] = -r$ (positive scaling dimensions correspond to relevant operators).

%%%%%%%%%%%%%%%%%%%%%%%%%%%%%%%%%%%%%%%%%%%%%%%%%%%%%

\subsection{SCR: Mapping to positive-$r$ spin-1 model and Kondo expansion in $1/J$}
\label{sec:scr}

\begin{figure}[t]
\includegraphics[width=0.42\textwidth]{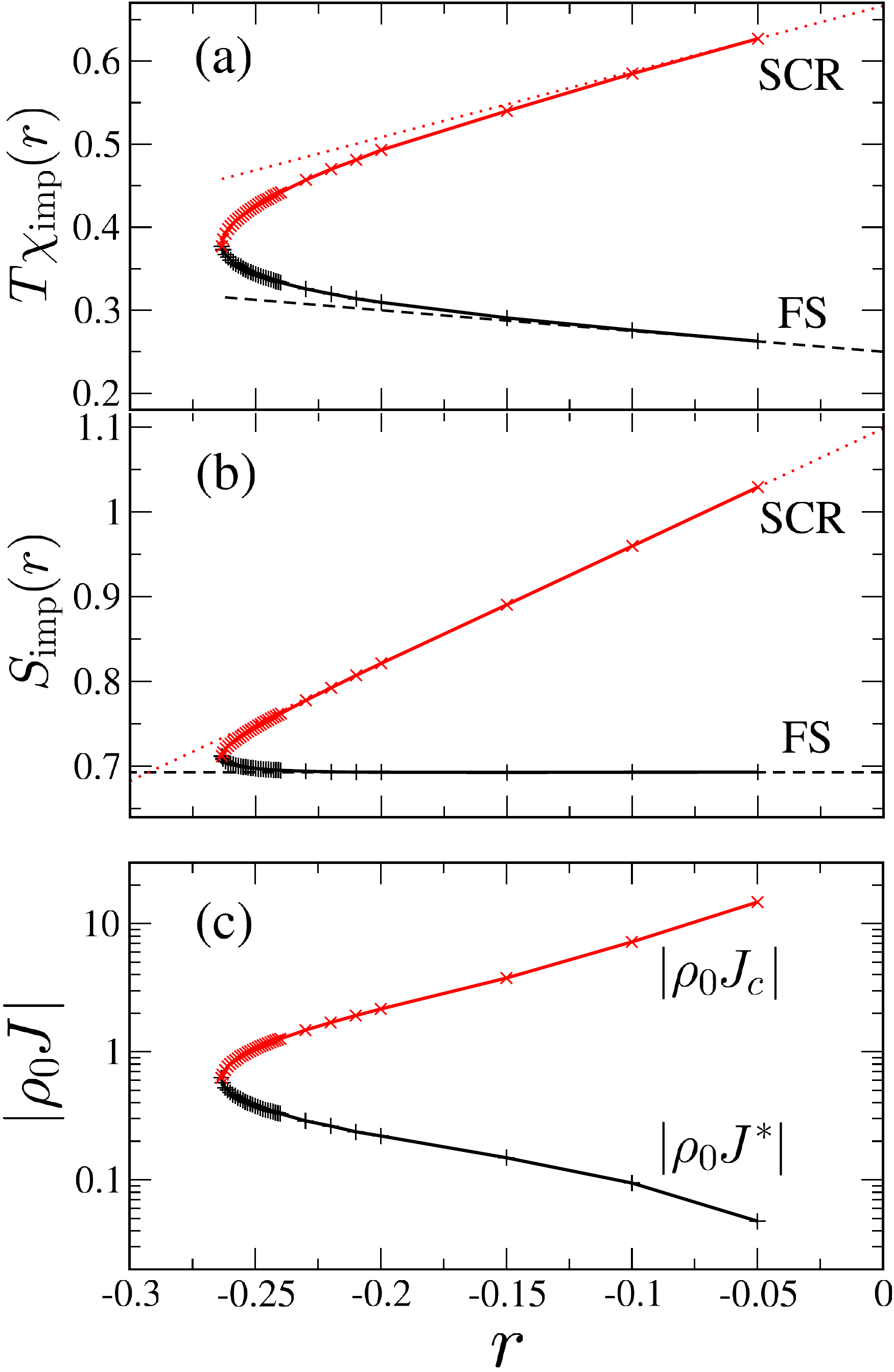}
\caption{
Universal $T\rightarrow 0$ values of (a) impurity magnetic moment, $\tchi$, and (b) impurity entropy,
$\s$, as function of the bath exponent $r$ for the fractional-spin fixed point (FS) and
the symmetric critical fixed point (SCR), obtained from NRG. Both FS and SCR exist for
$\bar{r}_{\rm max}<r<0$ with $\bar{r}_{\rm max}=-0.0264\pm 0.001$. The dashed lines show
the perturbative results in Eqs.~\eqref{sfs}, \eqref{tchifs} for FS; while dotted lines are Eqs.~\eqref{sscr}, \eqref{tchiscr} for SCR.
(c) Critical coupling $\rho_0 J_c$, for the TSC--FS transition controlled by SCR; and
renormalized intermediate coupling at the FS fixed point, $\rho_0 J^\ast$,
determined from a stationarity condition of the initial NRG flow.
} \label{fig:FS}
\end{figure}

The p-h symmetric critical (SCR) fixed point, controlling the transition between TSC and
FS, exists in the exponent range $\bar{r}_{\rm max}<r<0$. Numerical results [see also
Fig.~\ref{fig:FS}(c)] suggest that at SCR, the critical coupling $J_c \rightarrow -\infty$ diverges as $r\to 0^-$. This implies that SCR approaches TSC in this limit, so that
an expansion around TSC can be used to access SCR physics.

Indeed, such an expansion can be constructed using the insights gained in
Sec.~\ref{sec:chain}:
TSC, corresponding to $J=-\infty$, can be understood as a spin-1 impurity,
decoupled from a chain with $|\w|^{-r}$ DOS, Fig.~\ref{fig:chain}(f). Departing from
$J=-\infty$ allows (virtual) hopping processes between the first and second sites of the
original chain, which generate an effective exchange coupling
between the spin-1 object and the remainder of the chain.
Importantly, the effective coupling is {\em antiferromagnetic}, and scales as $t_{12}^2/|J|$
(where $t_{12}$ is the hopping matrix element connecting the first and second chain sites, and is on the order of the conduction bandwidth $D$).
Thus, for large ferromagnetic $|J|$, the negative-$r$ spin-$\tfrac{1}{2}$ Kondo model maps onto an
underscreened\cite{NB} spin-1 Kondo model --- but now with positive bath exponent $r'=-r$ and a
small antiferromagnetic coupling $J'\sim t_{12}^2/|J|$.

The QPT of this p-h symmetric spin-1 pseudogap model is amenable to a perturbative
treatment, which was worked out in Ref.~\onlinecite{serge05}. A double expansion in $r'$
and $J'$ yields the following critical-point properties: $\s= \ln 3 + \mathcal{O}(r'^3)$
and $\tchi = 2/3 - 2r'/3 + \mathcal{O}(r'^2)$ --- these are the thermodynamic
contributions of the spin-1 impurity relative to the $|\w|^{-r}$ bath. Note that $r=r'=0$
plays the role of a lower critical dimension.

To obtain the properties of SCR itself, we need to take into account the effect of the altered
chain length, which again leads to additional contributions corresponding to a resonant
level. Using $r'=-r$, we finally obtain the following results for SCR:
\begin{align}
\label{sscr}
\s^{\rm SCR} &= \ln 3 + 2r\ln 2 + \mathcal{O}(r^3) \,,\\
\tchi^{\rm SCR} &= \frac{2}{3} + \frac{r}{8} + \frac{2r}{3} + \mathcal{O}(r^2)\,.
\label{tchiscr}
\end{align}
These predictions are compared to NRG results as the dotted lines in Figs.~\ref{fig:FS}(a) and (b), and agree essentially perfectly at small $r$.
The correlation length exponent of SCR is\cite{serge05}
$1/\nu^{\rm SCR} = |r| + \mathcal{O}(r^2)$.
Although the anomalous susceptibility exponent was not previously calculated,
it can be obtained using the RG procedure of Ref.~\onlinecite{serge05}. We find it to be
$\eta^{\rm SCR}_\chi = r^2 + \mathcal{O}(r^3)$.

%%%%%%%%%%%%%%%%%%%%%%%%%%%%%%%%%%%%%%%%%%%%%%%%%%%%%

\subsection{AF-CR: Mapping to positive-$r$ spin-$\tfrac{1}{2}$ model and two expansions}

As we show below, the power-law Kondo model near its antiferromagnetic critical (AF-CR)
fixed point can be mapped onto an effective spin-$\tfrac{1}{2}$ p-h asymmetric pseudogap
Kondo model with bath exponent $r'=-r$. Since much of the relevant physics of the
power-law Kondo model in this regime can therefore be understood in terms of known
results for the pseudogap model, we review the latter here. Extensive discussions have
appeared elsewhere.\cite{GBI,fv04,mvrev,fvrop,DELlma}

\subsubsection{Review: RG flow of the pseudogap Kondo model}

\begin{figure}
\begin{center}
\includegraphics[width=0.38\textwidth]{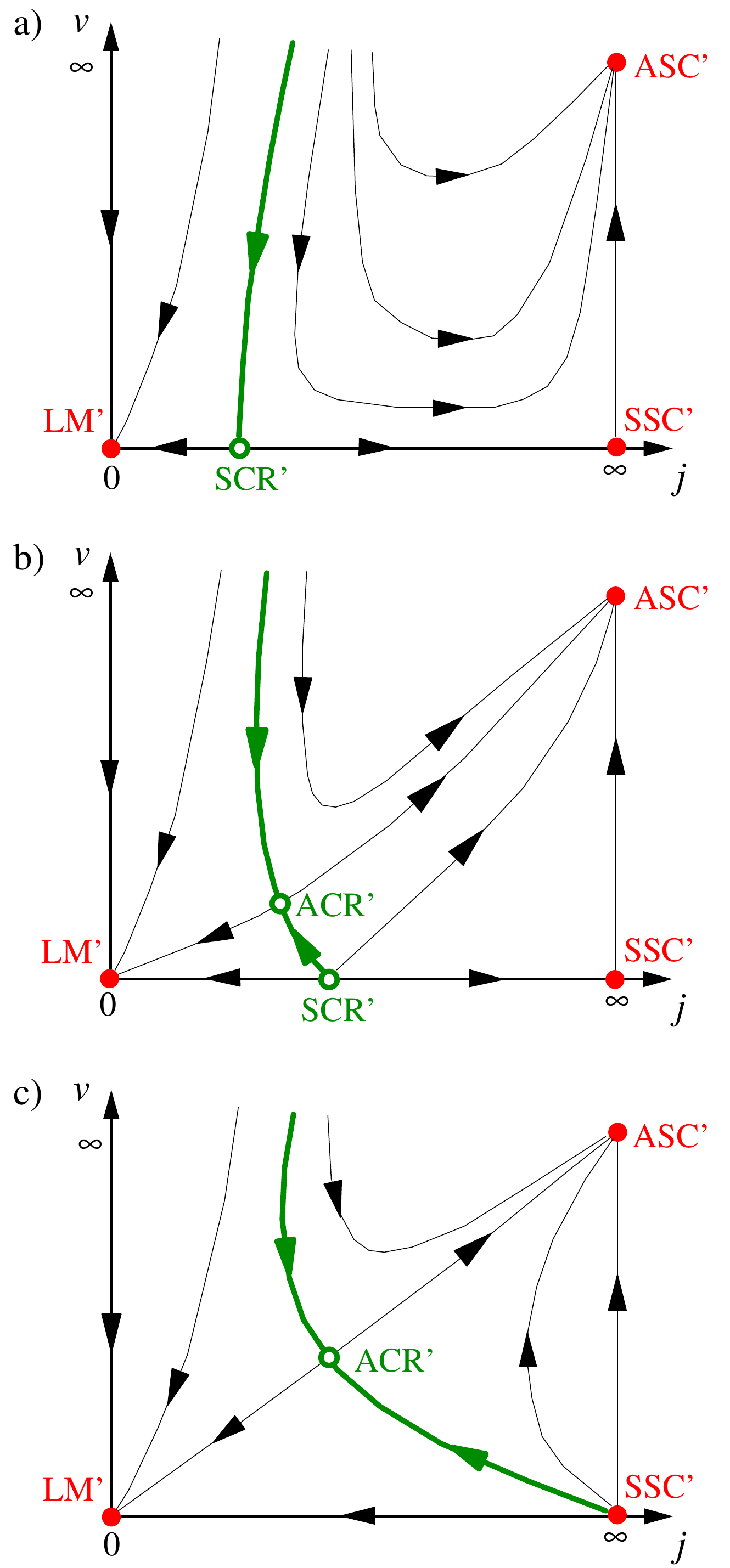}
\caption{
Schematic RG flow for the pseudogap Kondo model with DOS exponent $r'>0$ in the regimes
a) $0<r'<r^\ast$,
b) $r^\ast< r' <r_{\rm max}$,
c) $r' > r_{\rm max}$,
with $r^\ast = 0.375 \pm 0.002$ and $r_{\rm max} = \tfrac{1}{2}$.
As in Fig.~\ref{fig:flow}, the axes represent the renormalized Kondo coupling $j$
and potential scattering strength $v$. For details of the fixed points, see text.
}\label{fig:pgkm}
\end{center}
\end{figure}

As with the power-law Kondo model for $r<0$, the pseudogap Kondo model with $r'>0$
features three intervals of the bath exponent, characterized by qualitatively different RG flow, see
Fig.~\ref{fig:pgkm}. We restrict the following discussion to the relevant case of
antiferromagnetic Kondo coupling, and as before use primed labels for fixed points of
the pseudogap model, to distinguish them from the fixed points of the power-law Kondo model in Fig.~\ref{fig:flow}.

The flow diagram for $0<r'<r^\ast$ (with $r^\ast \approx 0.375$), is presented in
Fig.~\ref{fig:pgkm}(a). It contains two stable and two unstable
fixed points: LM$'$ corresponds to a free local moment and is equivalent to LM in
Sec.~\ref{sec:lm}. ASC$'$ denotes a p-h asymmetric strong-coupling fixed point with fully
screened impurity, $\s^{\rm ASC'}=0$ and $\tchi^{\rm ASC'}=0$.
SSC$'$ is equivalent to SSC in Sec.~\ref{sec:ssc}, and features a positive residual entropy
of $2r'\ln 2$. SSC$'$ is unstable towards ASC$'$ on breaking p-h symmetry.
A single (multi)critical p-h symmetric fixed point SCR$'$ separates LM$'$
from SSC$'$/ASC$'$.

For bath exponents $r^*<r'<r_{\rm{max}}=\frac{1}{2}$, the flow diagram changes to that in
Fig.~\ref{fig:pgkm}(b). Here, a p-h asymmetric critical (ACR$'$) fixed point splits off from
SCR$'$. The transition between LM$'$ and SSC$'$ tuned by $J$ at $V=0$ is still controlled by SCR$'$; but the transition between LM$'$ and ASC$'$ driven by p-h asymmetry is now controlled by ACR$'$.

Finally for $r'>r_{\rm{max}}$, Fig.~\ref{fig:pgkm}(c), SCR$'$ merges with SSC$'$. Thus, no Kondo screening can arise at p-h symmetry in this exponent range: the flow along the p-h symmetric axis
is towards LM$'$ only. The transition between LM$'$ and ASC$'$ on breaking p-h symmetry remains controlled by ACR$'$.

\subsubsection{AF-CR mapping}

Effective theories for AF-CR can be derived by noting that it approaches ALM in the
limit $r\to 0^-$, see Fig.~\ref{fig:phaseboundary}. Hence, an expansion around ALM, which
itself is characterized by $|V|=\infty$, appears appropriate.
Departing from $|V|=\infty$, (virtual) hopping processes between the first and second bath chain sites generate an effective exchange coupling betwen the impurity and the second chain site. This effective Kondo coupling, mediated by the (nearly) empty or doubly-occupied first chain site, is \emph{antiferromagnetic}, scaling as $J' \sim J t_{12}^2 /V^2$ ($t_{12}$ is again the tunneling matrix element between first and second chain sites). Since the DOS at the second chain site is $\rho(\omega) \propto|\omega|^{r'}$ (with $r'=-r$, see Sec.~\ref{sec:chain}), we obtain an effective p-h asymmetric pseudogap Kondo model. The transition between ALM and SSC is therefore equivalent to the transition between
LM$'$ and ASC$'$ in the pseudogap model with bath exponent $r'=-r$. This transition
is controlled by different fixed points depending on whether $r'\gtrless r^\ast$, see
Fig.~\ref{fig:pgkm}.

An alternative approach is to treat finite $V$ at AF-CR exactly according to
Eq.~\eqref{tdos}. On the lowest energy scales, this has the effect that the DOS
at the first chain site is converted from $\rho(\omega)\propto |\omega|^{r}$ to
$\rho(\omega)\propto |\omega|^{-r}$ (and with additional p-h asymmetry).
Again, a p-h asymmetric pseudogap Kondo model emerges.

\subsubsection{Kondo expansion}

For exponents $0<r'<r^*$ we know from Fig.~\ref{fig:pgkm} that the LM$'$-ASC$'$ transition
is controlled by SCR$'$. Its properties can be analyzed in a weak-coupling expansion in
the variables of the (effective) pseudogap Kondo problem, with coupling constants $J'$ and $V'$, the
latter parametrizing p-h asymmetry. To lowest order, the flow of the
dimensionless couplings $j'$ and $v'$ reads
\begin{eqnarray}\label{eq:pgflow}
\frac{d j'}{d \ln \Lambda}=j'\left(r'-j'\right)+\mathcal{O}\left(j'^3\right)\; ,\quad \frac{dv'}{d\ln \Lambda}=r'
v'\;,
\end{eqnarray}
with an unstable fixed point point predicted at $j'^\ast \simeq r'$ --- this is SCR$'$.
Observables have been calculated e.g. in Ref.~\onlinecite{kv04}; to convert them into
observables at AF-CR we again need to account for the altered chain length in the original power-law Kondo model. This is achieved simply by adding the constant offset between ALM and LM$'$, itself equivalent to an additional resonant level. Taken together we find
\begin{align}
\label{safcr1}
\s^{\rm AF-CR} &= \ln 2 + 2r \ln 2  +\mathcal{O}(r^3), \\
\tchi^{\rm AF-CR} &= \frac{1}{4}+\frac{3r}{8} +\mathcal{O}(r^2).
\label{tchiafcr1}
\end{align}
These perturbative results are compared with exact results from NRG as the dashed lines in Fig.~\ref{fig:afcr}, and agree essentially perfectly for $r\gg -r^\ast$ ($<0$).

The correlation length exponent follows as
$1/\nu^{\rm AF-CR} = |r| + \mathcal{O}(r^2)$, and the anomalous susceptibility exponent is found to be
$\eta^{\rm AF-CR}_\chi = r^2 + \mathcal{O}(r^3)$.

The mapping sketched above, together with a knowledge of the location of SCR$'$ at weak
coupling, allows us to infer the location of AF-CR. As discussed above, the potential scatterer can be incorporated into the bath part exactly, with the result that the DOS at the terminal chain site crosses over from $|\omega|^r$ to $|\omega|^{r'}$ behavior. The energy scale at which this happens is roughly set by $D'=\left(\frac{V \pi \rho_0}{D^r \cos(\pi r/2)} \right)^{-1/r}=\left(\frac{V \pi \rho_0}{D^r \cos(\pi r/2)} \right)^{1/r'}$. The running coupling at this scale is set by $j=\rho_0 J  \frac{D'^r}{D^r}$, which should be the critical coupling at AF-CR, $j^*$. From the (lowest order) RG scaling equation, Eq.~\eqref{eq:pgflow}, we anticipate that AF-CR arises at $j^*\simeq r'$, which yields a criterion for the phase boundary $(J_c,V_c)$ between ALM and SSC,
\begin{eqnarray}
V_c\simeq \frac{J_c/\pi }{-r},
\end{eqnarray}
which should be valid for small $r$. This result is consistent with NRG data, see Fig.~\ref{fig:phaseboundary}.

\subsubsection{Level-crossing expansion}

\begin{figure}
\includegraphics[width=0.43\textwidth]{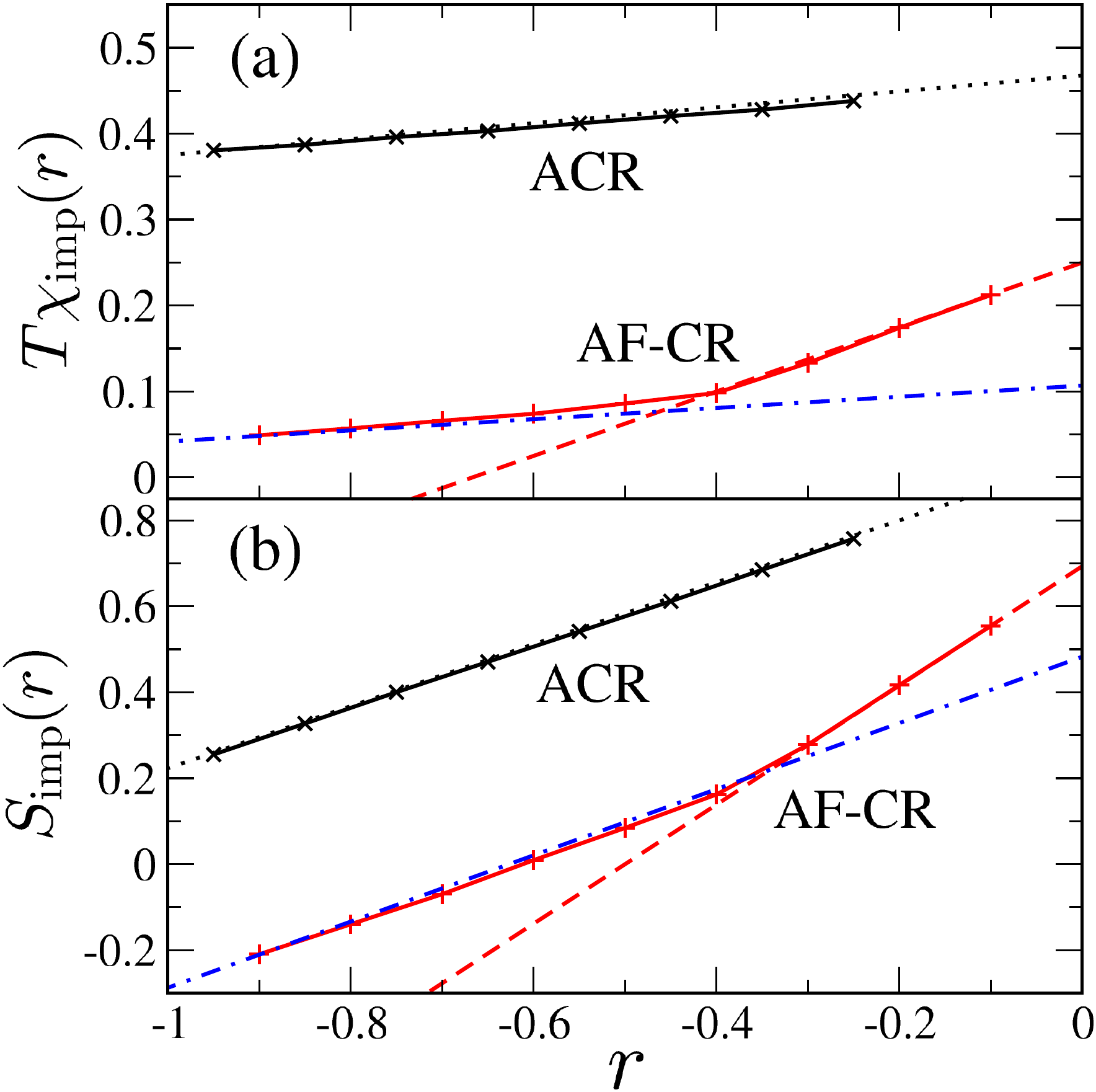}
\caption{
Universal $T\rightarrow 0$ values of (a) impurity magnetic moment, $\tchi$, and (b) impurity
entropy, $\s$, as function of the bath exponent $r$ for the critical fixed points
ACR and AF-CR, obtained using NRG (points and solid lines).
The dotted lines show the perturbative results for ACR, Eqs.~\eqref{sacr} and \eqref{tchiacr}; while the dashed and dot-dashed lines are for AF-CR, Eqs.~\eqref{safcr1}, \eqref{tchiafcr1}, \eqref{safcr2}, \eqref{tchiafcr2}. The properties of AF-CR display a pronounced change at $r \approx -0.4$, in
agreement with the prediction that AF-CR is equivalent to SCR$'$ for $r'= -r < r^\ast$, but to ACR$'$ for $r'=-r > r^\ast$ (with $r^\ast=0.375$).
}
\label{fig:afcr}
\end{figure}

For exponents $r'>r^*$, the LM$'$-ASC$'$ transition of the pseudogap Kondo model is
controlled by ACR$'$. It was realized in Ref.~\onlinecite{vf04} that the properties of
ACR$'$ can be captured in an expansion around the valence-fluctuation fixed point of a
maximally p-h asymmetric Anderson model, with the expansion being controlled in
$\bar{r}=1-r'$. The resulting observables were calculated in
Refs.~\onlinecite{vf04,fv04}.
Taking into account the offset corresponding to the altered chain length in the original $r<0$ model, we obtain
\begin{align}
\label{safcr2}
\s^{\rm AF-CR} &= \ln 3-\frac{8 \ln 2}{9}(1+r) + 2r \ln 2 +\mathcal{O}\left((1+r)^2\right), \\
\tchi^{\rm AF-CR} &= \frac{1}{6}- \left(\frac{1}{18}-\frac{\ln2}{27}\right) (1+r)+\frac{r}{8} +\mathcal{O}\left((1+r)^2\right).
\label{tchiafcr2}
\end{align}
Eqs.~\eqref{safcr2} and \eqref{tchiafcr2} are compared with exact results from NRG as the dot-dashed lines in Fig.~\ref{fig:afcr}, and again agree excellently for $r\ll -r^\ast$ ($>-1$).
In particular, we note that the change in the character of AF-CR
at $r'=-r \simeq r^\ast$ predicted by perturbative RG is nicely reflected in the numerical data.

The correlation length exponent is\cite{fv04}
$1/\nu^{\rm AF-CR} = |r| + \mathcal{O}((1+r)^2)$, and the anomalous susceptibility exponent follows as
$\eta^{\rm AF-CR}_\chi = 2(1+r)/3 + \mathcal{O}((1+r)^2)$.

%%%%%%%%%%%%%%%%%%%%%%%%%%%%%%%%%%%%%%%%%%%%%%%%%%%%%

\subsection{ACR: Mapping to positive-$r$ spin-1 model and level-crossing expansion}
\label{sec:acr}

The p-h asymmetric critical ACR fixed point controls the transition between TSC and ALM
in the power-law Kondo model. Remarkably, it is amenable to an expansion in a generalized
Anderson model, in analogy to the level-crossing expansions of
Refs.~\onlinecite{vf04,fv04,serge05}.

The effective theory can be constructed by noting that (i) TSC and ALM represent triplet
and doublet spin states, respectively, and (ii) both TSC and ALM involve, in the chain
representation of Sec.~\ref{sec:chain}, a bath chain with one site removed. This suggests
that the physics of ACR is captured by a minimal theory involving the crossing of a doublet and a triplet of levels, coupled to conduction electrons with DOS $\propto |\w|^{r'}$ (and $r'=-r$).

Such a fermionic field theory can be understood as a generalized infinite-$U$ Anderson
model, with maximal p-h asymmetry. This was considered first in Ref.~\onlinecite{serge05}
as a critical theory for the p-h asymmetric underscreened spin-1 pseudogap Kondo model.
As with the standard Anderson model,\cite{vf04,fv04} the
hybridization term is a marginal perturbation to the valence-fluctuation fixed point at
$r'=1$, which can thus be identified as an upper critical dimension.\cite{vf04}
This observation enables a perturbative treatment, technically performed as a double
expansion in $\bar{r} = 1-r'$ and the hybridization strength. Quoting the results from
Ref.~\onlinecite{serge05}, we have $\s= \ln 5 - 24\bar{r}\ln 2 /25 +
\mathcal{O}(\bar{r}^2)$ and $\tchi = 1/2 - (3-2\ln 2)\bar{r}/50 +
\mathcal{O}(\bar{r}^2)$. We note that the leading terms in $\s$ and $\tchi$ simply
correspond to the combined response of a spin-1 triplet and a spin-$\tfrac{1}{2}$ doublet.

The properties of ACR are now obtained by adding the resonant-level contribution as before, which
yield
\begin{align}
\label{sacr}
\s^{\rm ACR} &= \ln 5 + 2r\ln2 - \frac{24\ln 2}{25}(1+r) + \mathcal{O}((1+r)^2)\,,\\
\tchi^{\rm ACR} &= \frac{1}{2} + \frac{r}{8} - \frac{3-2\ln 2}{50}(1+r) + \mathcal{O}((1+r)^2)\,.
\label{tchiacr}
\end{align}
A comparison to exact NRG results is given in Fig.~\ref{fig:afcr} as the dotted lines, and agree extremely well over the entire exponent range.

The correlation length exponent of ACR is\cite{serge05}
$1/\nu^{\rm ACR} = |r| + \mathcal{O}((1+r)^2)$.
Along the lines of Ref.~\onlinecite{serge05}, one can also obtain the anomalous
susceptibility exponent, with the result
$\eta^{\rm SCR}_\chi = 2(1+r)/5 + \mathcal{O}((1+r)^2)$.

\subsection{$\bar{r}^\ast$ and $\bar{r}_{\rm max}$ from perturbation theory}
\label{critdim}

The mappings of Secs.~\ref{sec:fs}, \ref{sec:scr}, and \ref{sec:acr} allow one to infer
the separate existence of the fixed points FS, SCR (both for small negative $r$), and ACR
(for $r \gtrsim -1$). Remarkably, the RG analysis can even be used to predict the
emergence of the special exponent values $\bar{r}^\ast$ and $\bar{r}_{\rm max}$ where the
RG flow qualitatively changes, Fig.~\ref{fig:flow}. The basis for this is the fact that the
impurity entropy changes monotonically along the flow between two fixed points; this is a milder
version of the so-called $g$-theorem \cite{gtheorem} which states that the entropy should
decrease under RG flow (which strictly applies only to models with short-ranged interactions).

Applied to the pair of fixed points SCR and FS, with the flow topology as in
Fig.~\ref{fig:flow}a, this implies that both fixed points have to meet and disappear once
$\s^{\rm SCR}(r) = \s^{\rm FS}(r)$. Using the one-loop expressions in Eqs.~\eqref{sfs}
and \eqref{sscr} we find the estimate $\bar{r}_{\rm max} \approx -0.292$, which is to be compared 
with the exact value $\bar{r}_{\rm max} = -0.264 \pm 0.001$.

Similarly, the entropy matching condition for SCR and ACR using Eqs.~\eqref{sscr} and
\eqref{sacr} yields the estimate $\bar{r}^\ast \approx -0.232$, which is again close to the exact value
$\bar{r}^\ast = -0.245 \pm 0.005$. Interestingly, in the range $\bar{r}_{\rm max} < r <
\bar{r}^\ast$, we have $\s^{\rm ACR} > \s^{\rm SCR}$, such that the $g$-theorem is
violated. This is analogous to the situation for the positive-$r$ (pseudogap) Kondo model,\cite{fv04}
 itself related to the \emph{long-ranged} effective interactions in these models.

%%%%%%%%%%%%%%%%%%%%%%%%%%%%%%%%%%%%%%%%%%%%%%%%%%%%%
%%%%%%%%%%%%%%%%%%%%%%%%%%%%%%%%%%%%%%%%%%%%%%%%%%%%%
%%%%%%%%%%%%%%%%%%%%%%%%%%%%%%%%%%%%%%%%%%%%%%%%%%%%%

\section{Summary}

In this paper we revisited the rich physics of a Kondo impurity immersed in a fermionic
host with diverging power-law density of states near the Fermi level, $\rho(\w) \sim |\w|^r$, with exponent $r<0$.

This power-law model displays a number of stable phases and in total four non-trivial
intermediate-coupling fixed points, which we successfully described using different {\em
fermionic} critical field theories. We exploited various mappings to
effective models with positive bath exponent $(-r)$, which allowed us to make use of
existing field-theoretic results for variants of the pseudogap Kondo model. Altogether,
this demonstrates the remarkable versatility of the methods originally developed in
Refs.~\onlinecite{fv04,vf04}.

Our analytic results are in excellent agreement with exact numerical results obtained
from NRG: we conclude that the convergence radius of the epsilon expansion is sizeable here, but
in all cases restricted to $\epsilon<1$ (where $\epsilon$ is the deviation of the DOS
exponent from the critical ``dimension'' set by $r=0$ or $r=1$).

A detailed discussion of crossover functions, in particular in dynamical observables such as
the local spin susceptibility and the conduction-electron T matrix, is left for future
work.

%%%%%%%%%%%%%%%%%%%%%%%%%%%%%%%%%%%%%%%%%%%%%%%%%%%%%
%%%%%%%%%%%%%%%%%%%%%%%%%%%%%%%%%%%%%%%%%%%%%%%%%%%%%
%%%%%%%%%%%%%%%%%%%%%%%%%%%%%%%%%%%%%%%%%%%%%%%%%%%%%

\acknowledgments

This research was supported by the DFG through FOR 960 (AKM,RB,MV,LF), SFB 608 (AKM,RB), and FR2627/3-1 (LF), by GIF through grant G 1035-36.14/2009 (MV), by EPSRC through EP/I032487/1 (AKM), and the Institutional Strategy of the University of Cologne within the German Excellence Initiative (RB,LF).

%%%%%%%%%%%%%%%%%%%%%%%%%%%%%%%%%%%%%%%%%%%%%%%%%%%%%
%%%%%%%%%%%%%%%%%%%%%%%%%%%%%%%%%%%%%%%%%%%%%%%%%%%%%
%%%%%%%%%%%%%%%%%%%%%%%%%%%%%%%%%%%%%%%%%%%%%%%%%%%%%


\begin{thebibliography}{99}
\bibitem{error02}
The values of $T\chi_{\rm imp}$ quoted in Ref.~\onlinecite{vb02} for $r<0$ were
incorrect at low temperatures, due to an erroneous subtraction of the bulk contribution to $T\chi$. In particular, high-accuracy NRG calculations with enforced symmetries are required due to numerical instabilities for ferromagnetic Kondo coupling $J<0$.
\end{thebibliography}
\end{document}